\documentclass[fleqn,usenatbib,useAMS]{mnras}
\usepackage{lmodern}
\usepackage{bm}

\usepackage{graphicx,epsfig}
\usepackage{url}

\usepackage{amsmath}
\usepackage{amssymb}

\usepackage{threeparttable}

\usepackage{ifsym}

\usepackage{xcolor}

\usepackage{longtable}

\usepackage{tabularray}

\usepackage{supertabular,booktabs}

\usepackage{float}

\newcommand{\kms}{\,km\,s$^{-1}$}
\newcommand{\msun}{M$_{\odot}$}
\usepackage{bm}

\newcommand{\teff}{\mbox{$T_{\mathrm{eff}}$}}
\newcommand{\teffp}{\mbox{$T_{\mathrm{eff,1}}$}}
\newcommand{\teffs}{\mbox{$T_{\mathrm{eff,2}}$}}

\newcommand{\loggp}{\ensuremath{\log g_{\mathrm 1}}}
\newcommand{\loggs}{\ensuremath{\log g_{\mathrm 2}}}

\newcommand{\rsun}{R\ensuremath{_\odot}}

\newcommand{\mprim}{\ensuremath{M_{\mathrm {1}}}}
\newcommand{\msec}{\ensuremath{M_{\mathrm {2}}}}
\newcommand{\rprim}{\ensuremath{R_{\mathrm {1}}}}
\newcommand{\rsec}{\ensuremath{R_{\mathrm {2}}}}

\newcommand{\degree}{\mbox{\ensuremath{^\circ}}}

\title[The intermediate-mass PMS binary $\theta^1$~Ori~E]{Medium-resolution spectroscopic study of the intermediate-mass pre-main sequence binary $\theta^1$~Ori~E }

\author[Costero, Echevarr\'ia, G{\'o}mez Maqueo Chew, Ruelas-Mayorga, and S\'anchez]{Rafael Costero$^1$\thanks{In active retirement; E-mail: costero@astro.unam.mx }, Juan Echevarr\'ia$^1$,  Yilen G{\'o}mez Maqueo Chew$^1$, 
\newauthor
Alex Ruelas-Mayorga$^1$ and Leonardo J. S\'anchez$^1$\\$^1$Instituto de Astronom{\'\i}a, Universidad Nacional Aut\'onoma de M\'exico, Apartado Postal 70-264, 04510 M\'exico, CDMX, M\'exico \\
}

\date{Accepted 2025 October 31. Received 2025 October 08.}
\pubyear{2025}

\begin{document}


\maketitle


\begin{abstract}
$\theta^1$~Ori~E is a very young and relatively massive pre-main sequence (PMS) spectroscopic and eclipsing binary with nearly identical components. We analyze Échelle spectra of the system obtained over fifteen years and report 91 radial velocities measured from cross-correlating the observations with a suitable synthetic spectrum. The spectra of individual binary components are indistinguishable from each other, with a composite spectral type around G4\,III. 
The projected equatorial velocity is estimated to be $v\,sin\,i=32\pm 3~km~s^{-1}$, consistent with rotational synchronization.
We find that the circular orbit has $P_{\rm orb} = 9.89522 \pm 0.00003~d$, $K_1 = 83.36 \pm 0.29~km~s^{-1}$, $K_2 = 84.57 \pm 0.28~km~s^{-1}$, and $asini = 32.84\pm0.08\ R_\odot$. The mass ratio is $ q = 0.9856 \pm 0.0047$, indicating nearly identical but significantly different masses. The systemic velocity of the binary, $\gamma = 29.7 \pm 0.2~km~s^{-1}$, is similar to that of other Trapezium members. 
Using Spitzer light curves and our results, we derive $M_1 = 2.755\pm0.043\ M_{\odot}$, $M_2 = 2.720\pm0.043\ M_{\odot}$, $R_1=6.26\pm0.31R_{\odot}$ and $R_2=6.25\pm0.30R_{\odot}$. 
Together with our estimate of the effective temperature, $T_{eff}=5150\pm200\ K$, a bolometric luminosity of $28.8\pm4.6\ L_{\odot}$ is derived for each component. Compared to evolutionary models of PMS stars, the binary age turns out to be less than or equal to $\sim 10^5$ years. Its components are probably the most massive stars known with masses determined with precision better than 2 percent, with both being PMS stars.
\end{abstract}




\maketitle


\section{Introduction and previous information}
\label{sec:intro}

$\theta^1$~Ori~E, a member of Orion Trapezium, has many other acronyms used in the literature, including Brun 584, ADS 4186 E, Parenago 1864, NSV~2291 and HC344. Identification charts for this and other members of the Trapezium can be found elsewhere \citep[e.g.][]{wolf94, herfin06}. 

In the \citet{par54} compilation of the Orion Nebula Cluster (ONC), the star -- listed on page 342 -- is assigned photographic and photovisual magnitudes of 11.7 and 11.4, respectively, and a G + B5--B8 spectral type. This spectral classification is probably based on a note in a paper by \citet{her50} on emission stars in the ONC; \citet{herfin06} later recognized that the B5--B8 component was light contamination from the nearby Component A of the Trapezium.
Indeed, $\theta^1$~Ori~E is located 4.5\,arcsec north of the Trapezium Component A -- a B\,0.5\,V spectroscopic and eclipsing binary (V1016 Ori), about 3.5 visual magnitudes brighter than E -- and 6.2 arcsec south-west of Component B, an early-B-type eclipsing and spectroscopic binary star (BM Ori) a little more than 2 visual magnitudes brighter than E. Such stellar agglomeration, aggravated by the very bright emission nebula surrounding the trapezium, has hindered all reliable information about its weaker components, gathered before the end of last century.

In fact, since shortly after its discovery in 1826 -- attributed to Wilhem Struve by \citet{webb1859} -- $\theta^1$ Ori E has been suspected (and only recently confirmed) to vary in brightness, including the possible gradual secular brightening of the star, as suggested by \citet{gled1880}. Further reports of possible variability of the star in the optical range \citep[e.g.,][]{{walker77, FaG78}} led to its inclusion in the Catalog of Suspected Variable Stars under number 2291 \citep{kak82}, where a variability amplitude of 0.41\,mag is assigned to it. However, we know of only two papers in which the Trapezium Component E is undoubtedly found to be variable at optical wavelengths: \citet{wolf94} and \citet{mor12}.
In the first one, {\sc PSF} photometry of the Orion Trapezium stars is conducted to obtain the light-curve of the eclipsing component B, using the Str\"omgren {\it uvby} and Cousins {\it R, I} photometric systems; there, it is stated that ``definite night-to-night variations of several tenths of magnitude" are observed in $\theta^1$ Ori~E. Regrettably, no numerical information, is given there.
In the second paper, \citet{mor12} perform a careful search for eclipsing binaries among ONC members using the {\it Spitzer} telescope survey of the cluster at 3.6\,$\micron$ and 4.5\,$\micron$; they find that Component E is a grazing eclipsing star and note that it varied by 0.064\,mag in the 4.5 $\micron$ band between the two observation epochs, separated by 13\,months. 

Variability of $\theta^1$~Ori~E in other wavelength ranges has been firmly identified. The star, discovered to be a radio source by \citet{gar87, gar89} (listed there as object 25) was later found by \citet{fea93} to vary at 2\,cm and 6\,cm by about 50 percent. However, \citet{zea04} do not register variations of more than 20 percent at 3.7\,cm. To explain these contradictory results, the latter speculates that this and other radio sources in the Trapezium are a combination of a non-thermal stellar irregular variable emitters and a constant thermal source. 
In contrast, this binary is a very bright \citep{kea82} and strongly variable X-ray emitter \citep [see Table 7 in][]{gea05}; in fact, it is the second brightest X-Ray source in the Trapezium after Component C (the main ionization source of the emission nebula). Recently, a thorough photometric and spectroscopic X-Ray survey of the ONC was released by \citet{shulz24}. There, the authors present a whole section dedicated to $\theta^1$~Ori~E, including high resolution spectra and a cumulative light curve of the binary. Intense flare-like variability of nearly one order of magnitude, on timescales of thousands of seconds, is interpreted as magnetic reconnection events resulting from stellar coronal flares from Component E. 

Other previous findings about $\theta^1$~Ori~E, relevant to the present work, are: 
\begin{enumerate}
\item Based on historical measurements of the relative position of components A and E, which span more than 170 years, \citet{apw74} concluded that component E escapes the Trapezium system with a transverse velocity of about $5\,km~s^{-1}$ with respect to Component A. The lack of radial velocity measurements of $\theta^1$~Ori~E was precisely the reason why, in 2004, Prof. Arcadio Poveda suggested that we obtain some measures as part of other observing programs. The result of \citet{apw74} was later confirmed by \citet{oli13} who, in addition to historical measurements, used diffracto-astrometric determinations based on HST WFP Camera 2 archival images, as well as our preliminary results for the systemic radial velocity of the binary \citep{cos08}.

\item $\theta^1$\,Ori\,E was first reported to be a double-lined PMS spectroscopic binary with nearly identical components by \citet{cea06}. Soon after, this was confirmed by \citet{herfin06}. The latter authors used ten high-resolution and signal-to-noise spectra to obtain, for the first time, orbital parameters of the binary.

\item \citet{Allenetal2015,Allenetal2017} concluded that Component B in the Orion Trapezium (itself a minicluster with at least 6 stellar members) and the Trapezium as a whole have very short dynamical lives of the order of $10^4$ years. 

\item An upper limit of $0.7\,mJy$ at $\lambda 850\,\mu m$ was obtained by \citet{Eisner2018} for the emission of circumstellar material surrounding this binary. The absence of a significant circumbinary disk may indicate that the stellar system is in a rather evolved stage or that the nearby massive stars have removed the material in which it might have been embedded. 
\end{enumerate}

In this paper, we present the radial velocities of the binary components of $\theta^1$\,Ori\,E measured in 91 spectra, as well as the results derived from them and from 18 spectra obtained near conjunction, all acquired during a period of more than 15\,years. In Section \ref{sec:Obs} our observations are described, and in Section \ref{sec:ResDisc} we present our results for the orbital parameters, describe the binary spectra, derive the physical characteristics of its almost identical stars, estimate their composite temperature and rotational velocity, and inquire into the possibility of detecting the Rossiter-McLaughlin (RM) effect in the spectra obtained during the eclipsing phases. Finally, in Section \ref{sec:conclusions}, we discuss additional findings that we obtain from previously published information, such as its infrared light curve and proper motions as derived from Gaia or \rm{VLBA} observations; and discuss its position with respect to theoretical evolutionary tracks of PMS. A comprehensive summary is given in Section \ref{sec:sum}.


\section{Observations}
\label{sec:Obs}

We observed $\theta^1$, Ori, E between October 2004 and January 2020, with the Échelle spectrograph attached to the 2.1m Telescope of the Observatorio Astron\'omico Nacional at San Pedro M\'artir, Baja California, M\'exico. Several CCD detectors were used, ranging from 1024$\times$1024, 24\,\micron\, pixel CCD detectors to a back-illuminated 2048$\times$2048, 13.5\,$\micron$\,pixel. The spectral resolution was $R \sim 12,000$. All observations were carried out with the 300\,l/mm cross-dispersor, blazed at around $\lambda$5500\,\AA~in its first dispersion order.  Exposure times per frame ranged from 600 to 1200 seconds, depending on observing conditions and they are listed in the second column of Tables~\ref{tab:RadVelTab} and \ref{tab:LogConj}. 

Data were reduced using standard {\sc iraf} routines for Échelle CCD spectroscopy, which included bias correction, tracing and extraction of the spectrum, and wavelength calibration using bracketing ThAr lamps for each group of consecutive spectra. A log of all the spectra obtained when the binary was not near conjunction is presented in Table~\ref{tab:RadVelTab}; those near conjunction are listed in Table~\ref{tab:LogConj}. 


\section{Results}
\label{sec:ResDisc}

   %
   
\subsection{Spectral type and peculiarities}
\label{sec:Spectype}

The spectral lines of the two stellar components of $\theta^1$\,Ori\,E, when clearly separated, are nearly identical to each other; their mutual blending complicates their adequate identification. Hence, it is appropriate to use the composite spectrum of the binary obtained near conjunction, such as those listed in Table~\ref{tab:LogConj}, to identify the lines and classify the composite spectrum. After close inspection of these spectra and considering the information included in this paper and that by \citet{mor12}, we conclude that the composite spectrum is that of a G4\,III star, in good agreement with the spectral type estimated by \citet{herfin06} for this star (G5\,III). Other spectral classifications in the literature are G2\,IV by \citet{cos08} and G4-K3 obtained from low-dispersion near-infrared spectra by \citet{Luhman2000}. The spectral type F8Vn, attributed to H. Abt by \citet{kea82}, was not confirmed by Professor Abt in a personal communication with one of us (RC). 

As already noted by \citet{herfin06} and \citet{cos08}, the two notable peculiarities of the spectrum that are strongly indicative of the PMS nature of this double-lined spectroscopic binary are:
\begin{enumerate}
    \item The very strong lithium resonant doublet at $\lambda 6708$ \AA, with an equivalent width of about $0.36$\,\AA \ (probably veiled) in each component of the binary system. 
    \item The Ca II emission in the core of the resonant $\lambda 3933$\,\AA \ absorption line, present in both components.
\end{enumerate}

A more careful inspection of the composite spectrum of the binary reveals that certain metallic lines, specifically those of Cr\,I, seem to be weaker, relative to those of Fe\,I, than expected in stars with solar abundances and, hence, predict lower temperatures than those derived from the ratio of lines of the same species (see Section~\ref{sec:Temp}). This possibly means that certain metals may be underabundant in $\theta^1$\,Ori\,E. However, a detailed analysis to determine and explore this possible underabundance of some elements is beyond the scope of this work. 


\subsection{Radial velocities}
\label{sec:RadialVel}

To derive the radial velocities of both components of $\theta^1$~Ori~E, after cleaning each wavelength-calibrated spectrum of obvious cosmic-ray hits, we averaged the spectra in groups, usually of three, flanked by wavelength calibration spectra of a Th-Ar lamp. The spectra within each group are close enough in time so that the corresponding change in their observed radial velocity is smaller than the instrumental error. 

This resulted in 109 spectra, 91 of which are suitable for determining the orbital parameters of the binary, because the spectra of both components are sufficiently separated in wavelength. The dates of acquisition, exposure times, and the radial velocity measurements of these 91 spectra are listed in Table~\ref{tab:RadVelTab}. The other 18 spectra, listed in Table~\ref{tab:LogConj}, are very close to conjunction, and consequently, the spectra of both components are strongly blended. They were obtained less than 0.040 in phase (about 9.5\,hours) away from conjunction and, consequently, during the nearly grazing eclipse discovered by \citet{mor12} in this star.

The derived radial velocities, listed in both tables, were obtained as follows: after excising the few and weak nebular emission lines present in the spectral range between $\lambda$5017\,\AA\, and $\lambda$5667\,\AA  \  (as well as the telluric [O\,I]\,$\lambda$5577\,\AA \ line when noticeable), the five consecutive \'Echelle orders that cover that range were first normalized to their apparent continuum and then cut near their respective free spectral ranges. Finally, the resulting segments were added to form single one-dimensional spectra.
This spectral interval was chosen because: a) many strong metallic lines are present in the spectrum of $\theta^1$~Ori~E; b) the response of the cross-disperser in the \'Echelle spectrograph is near its maximum; c) the contaminating spectrum of $\theta^1$~Ori A presents few and very weak absorption lines; d) telluric absorption lines are scarce and very weak, and e) there are few detectable emission nebular lines \citep[for a comprehensible list of emission lines in the Orion Nebula see][]{esal04}.

The radial velocities of both binary components were obtained by the cross-correlation of the 91 spectra in Table~\ref{tab:RadVelTab} with a suitable synthetic spectrum (see below), using the {\sc iraf} routine {\sc fxcor}. The two sets of absorption lines of the double lined spectroscopic variable were always measured simultaneously by means of the {\sc deblend} option in the {\sc fxcor} routine, even around quadrature, when the two line systems are well separated. In doing so, we have assumed that the correlation width is the same for both stars in any particular spectrum. This assumption is based on the stars being nearly indistinguishable from each other and on the binary being in rotational synchronization, as proved in Section \ref{sec:RotatVel} below.

The 18 spectra listed in Table~\ref{tab:LogConj} were originally acquired under the possibility of the existence of a sizable eclipse \citep[]{herfin06} that would produce a detectable Rossiter-McLaughlin (RM) effect on their measured radial velocities. If so, several physical parameters of the orbit and of the component stars could be inferred (see Section ~\ref{sec:RMEffect}). The radial velocity measured in these spectra was obtained by adjusting a single Gaussian in their cross-correlation with the same synthetic spectrum used in the double-lined spectra.

The synthetic spectrum used was chosen from a grid calculated for temperatures between 6000\,$K$ and 5000\,$K$ (in steps of 250\,$K$) and $log\,g$ between 2.5 and 4.5 in steps of $0.5\,dex$. These spectra were kindly produced by Prof. Leonid N. Georgiev (shortly before his passing) using the {\sc atlas9} stellar atmosphere models by \citet{cas04} and calculated applying the {\sc synthe} program developed by R.L.\ Kurucz. In doing so, a rotational velocity of $30\,km~s^{-1}$, solar abundances, and a degraded spectral resolution of $20\,km~s^{-1}$ were used. From these synthetic spectra we selected those computed for $log\,g=3.5$ \citep[the closest value to $log\,g=3.28\pm 0.04$, derivable from the results of][assuming both components have identical radii]{mor12} to be initially cross-correlated with the spectra of our object. From these, we then chose the one calculated for $T=5250\,K$ because it was the one that produced the highest correlation height.
 
All the radial velocities obtained were then corrected for zero-point shifts, as described in detail by \citet{cos21} and summarized in what follows: the zero-point corrections were estimated from the difference between the average of the measured radial velocity of strong, non-saturated nebular lines -- namely HeI\,$\lambda\lambda$ 4171 and 5875, H$\beta$, and [OIII]\,$\lambda$4959\,\AA -- and that measured by \citet{cas88} for [OIII]\,$\lambda$5007\,\AA  \, around $\theta^1$\,Ori\,E. We estimate the velocity of the [OIII] lines at the location of $\theta^1$\,Ori\,E to be $18.7\,km~s^{-1}$. In this case, this velocity is relatively uncertain (about $2\,km~s^{-1}$) because the nebular velocity varies strongly around that region \citep[see][]{cas88}. The zero-point corrections calculated rarely exceeded $6\,km~s^{-1}$ and their average along the entire observation lapse is nearly zero, making them a reasonable choice of zero-point correction for our observations. 

The resulting radial velocities, together with their corresponding errors, are listed in Tables~\ref{tab:RadVelTab} and \ref{tab:LogConj}. The Heliocentric Julian Date at mid-exposure and the number of the averaged spectra times the exposure time of each are given in the first two columns of both tables.

\begin{table}\centering
\caption{Log of $\theta^1$\,Ori\,E spectra and measured radial velocities}
\label{tab:RadVelTab}
\begin{tabular}{ccccc}
\toprule
HJD & $t_{\rm exp}$ & Primary & Secondary & Phase \\
$-2450000$ & (s) & (km~s$^{-1}$) & (km~s$^{-1}$) & \\
\hline
3284.97657 & 1x600  & -21.6 $\pm$ 4.1 &  78.2 $\pm$ 4.2 & 0.397 \\
3295.04042 & 2x600  & -15.6 $\pm$ 4.0 &  71.3 $\pm$ 4.0 & 0.414 \\
3349.88589 & 1x600  &  49.1 $\pm$ 4.9 &   9.3 $\pm$ 5.1 & 0.957 \\
3351.86195 & 3x600  & -39.7 $\pm$ 4.6 &  99.1 $\pm$ 4.7 & 0.157 \\
3352.85855 & 3x600  & -54.1 $\pm$ 5.2 & 115.8 $\pm$ 5.1 & 0.257 \\
3353.84004 & 3x600  & -34.7 $\pm$ 4.4 &  96.1 $\pm$ 4.8 & 0.356 \\
3743.79639 & 2x600  & 113.7 $\pm$ 4.6 & -57.0 $\pm$ 4.3 & 0.765 \\
3743.82948 & 3x300  & 113.0 $\pm$ 4.9 & -55.5 $\pm$ 4.9 & 0.768 \\
3743.85376 & 3x600  & 113.5 $\pm$ 4.6 & -54.9 $\pm$ 4.4 & 0.771 \\
3743.89958 & 3x600  & 111.6 $\pm$ 4.9 & -56.3 $\pm$ 4.6 & 0.775 \\
3744.75601 & 2x600  &  92.1 $\pm$ 4.2 & -35.1 $\pm$ 3.6 & 0.862 \\
3744.84105 & 1x600  &  90.7 $\pm$ 4.2 & -29.8 $\pm$ 3.5 & 0.871 \\
3745.63934 & 3x600  &  60.6 $\pm$ 4.5 &   8.3 $\pm$ 3.6 & 0.951 \\
3745.67939 & 3x600  &  52.4 $\pm$ 4.7 &   5.4 $\pm$ 3.6 & 0.955 \\
3745.71866 & 3x600  &  50.6 $\pm$ 4.9 &   8.3 $\pm$ 3.6 & 0.959 \\
3746.60088 & 3x600  &  -0.6 $\pm$ 4.1 &  52.0 $\pm$ 3.7 & 0.048 \\
3746.65782 & 3x600  &   0.2 $\pm$ 3.8 &  58.0 $\pm$ 4.1 & 0.054 \\
3747.63624 & 6x600  & -37.9 $\pm$ 4.1 &  98.7 $\pm$ 4.2 & 0.153 \\
3747.78561 & 3x600  & -40.9 $\pm$ 4.4 & 104.3 $\pm$ 4.2 & 0.168 \\
3747.87593 & 3x600  & -43.6 $\pm$ 5.2 & 106.1 $\pm$ 4.7 & 0.177 \\
3748.62612 & 3x600  & -56.6 $\pm$ 4.7 & 112.1 $\pm$ 4.4 & 0.253 \\
3748.68132 & 3x600  & -55.7 $\pm$ 4.7 & 113.3 $\pm$ 4.3 & 0.259 \\
3750.62133 & 3x600  &   7.9 $\pm$ 4.2 &  55.4 $\pm$ 4.7 & 0.455 \\
3751.62061 & 4x600  &  61.2 $\pm$ 5.9 &   1.4 $\pm$ 5.9 & 0.556 \\
3751.85749 & 1x600  &  76.2 $\pm$ 6.3 &  -9.5 $\pm$ 6.6 & 0.580 \\
3752.61545 & 3x600  & 100.1 $\pm$ 4.9 & -38.0 $\pm$ 4.5 & 0.656 \\
3752.87657 & 3x600  & 106.7 $\pm$ 5.0 & -46.7 $\pm$ 4.6 & 0.683 \\
3791.63133 & 2x600  &  77.6 $\pm$ 4.5 & -18.4 $\pm$ 4.0 & 0.599 \\
3807.69462 & 3x900  & -51.7 $\pm$ 5.1 & 114.7 $\pm$ 4.8 & 0.223 \\
3808.64557 & 3x900  & -45.9 $\pm$ 4.9 & 107.0 $\pm$ 4.5 & 0.319 \\
4076.76736 & 1x1200 & -12.2 $\pm$ 4.1 &  73.8 $\pm$ 4.2 & 0.415 \\
4078.70560 & 1x1200 &  80.0 $\pm$ 4.1 & -25.0 $\pm$ 3.9 & 0.611 \\
4078.80481 & 1x1200 &  84.7 $\pm$ 4.4 & -27.6 $\pm$ 3.8 & 0.621 \\
4078.92758 & 1x1200 &  90.4 $\pm$ 5.0 & -31.2 $\pm$ 4.3 & 0.633 \\
4078.96329 & 1x1200 &  90.7 $\pm$ 4.4 & -32.6 $\pm$ 4.1 & 0.637 \\
4079.84588 & 1x1200 & 113.3 $\pm$ 5.2 & -52.7 $\pm$ 5.2 & 0.726 \\
4080.83999 & 1x1200 & 105.2 $\pm$ 4.6 & -44.0 $\pm$ 4.6 & 0.826 \\
4081.74079 & 1x1200 &  71.8 $\pm$ 4.2 & -12.2 $\pm$ 4.1 & 0.917 \\
4352.00905 & 3x600  & -54.5 $\pm$ 5.1 & 112.8 $\pm$ 4.5 & 0.230 \\
4353.01959 & 2x600  & -40.7 $\pm$ 4.9 & 104.8 $\pm$ 4.5 & 0.332 \\
4426.87823 & 1x1200 & 110.8 $\pm$ 5.1 & -50.5 $\pm$ 5.0 & 0.797 \\
4427.89811 & 1x1200 &  77.5 $\pm$ 4.6 & -21.0 $\pm$ 4.0 & 0.900 \\
4476.78216 & 3x900  & 100.4 $\pm$ 4.9 & -40.7 $\pm$ 4.4 & 0.840 \\
4477.88541 & 1x900  &  55.9 $\pm$ 4.7 &   4.8 $\pm$ 4.5 & 0.951 \\
4480.64940 & 1x900  & -52.4 $\pm$ 4.9 & 115.9 $\pm$ 5.4 & 0.231 \\
4481.63425 & 1x900  & -43.5 $\pm$ 4.4 & 104.0 $\pm$ 4.8 & 0.330 \\ 

\toprule
\end{tabular}
\end{table}

\begin{table}\centering
\begin{flushleft}
{\bf Table 1.} Continued
\end{flushleft}
\vspace{0.2cm}
\begin{tabular}{ccccc}
\toprule
HJD & $t_{\rm exp}$ & Primary & Secondary & Phase \\
$-2450000$ & (s) & (km~s$^{-1}$) & (km~s$^{-1}$) & \\
\hline

4484.66408 & 1x900  &  91.5 $\pm$ 4.6 & -33.0 $\pm$ 4.1 & 0.636 \\
4484.81222 & 1x900  &  98.7 $\pm$ 5.3 & -37.6 $\pm$ 5.3 & 0.651 \\
4871.60723 & 1x900  & 118.4 $\pm$ 5.0 & -49.1 $\pm$ 5.0 & 0.740 \\
4873.64327 & 2x900  &  59.2 $\pm$ 4.2 &   3.1 $\pm$ 4.5 & 0.946 \\
4875.62996 & 2x900  & -32.7 $\pm$ 4.3 &  99.3 $\pm$ 4.8 & 0.147 \\
4876.61833 & 1x900  & -53.7 $\pm$ 4.9 & 117.1 $\pm$ 5.6 & 0.247 \\
5203.85835 & 3x900  & -48.4 $\pm$ 4.7 & 106.0 $\pm$ 5.3 & 0.317 \\
5204.76366 & 3x900  & -15.2 $\pm$ 3.6 &  76.1 $\pm$ 4.6 & 0.409 \\
5207.81097 & 3x900  & 108.8 $\pm$ 4.7 & -57.6 $\pm$ 5.9 & 0.717 \\
5208.78686 & 3x900  & 104.9 $\pm$ 4.9 & -48.4 $\pm$ 6.1 & 0.815 \\
5208.82344 & 3x900  & 105.4 $\pm$ 4.9 & -48.0 $\pm$ 5.9 & 0.819 \\
5208.86070 & 3x900  & 102.9 $\pm$ 5.0 & -47.9 $\pm$ 6.4 & 0.823 \\
5208.92353 & 2x900  & 100.5 $\pm$ 5.0 & -48.4 $\pm$ 6.1 & 0.829 \\
5209.74213 & 3x900  &  69.9 $\pm$ 4.4 & -18.9 $\pm$ 4.5 & 0.912 \\
5530.89636 & 3x900  & -30.3 $\pm$ 5.6 &  90.3 $\pm$ 5.6 & 0.367 \\
5533.02627 & 1x900  &  67.4 $\pm$ 4.2 & -16.2 $\pm$ 4.4 & 0.583 \\
5535.86687 & 1x900  &  90.4 $\pm$ 4.2 & -29.6 $\pm$ 4.3 & 0.870 \\
5574.89206 & 2x600  & 105.3 $\pm$ 5.4 & -47.3 $\pm$ 6.0 & 0.814 \\
5578.71334 & 1x900  & -50.0 $\pm$ 4.7 & 110.4 $\pm$ 4.7 & 0.200 \\
5934.81580 & 1x900  & -44.9 $\pm$ 5.2 & 107.2 $\pm$ 5.7 & 0.187 \\
5935.74987 & 1x900  & -51.4 $\pm$ 5.8 & 113.0 $\pm$ 5.5 & 0.281 \\
5936.71994 & 1x900  & -30.0 $\pm$ 4.9 &  83.8 $\pm$ 4.7 & 0.379 \\
5970.64518 & 1x900  & 107.4 $\pm$ 5.3 & -52.8 $\pm$ 5.6 & 0.808 \\
5971.66179 & 2x900  &  75.9 $\pm$ 5.6 & -13.4 $\pm$ 6.3 & 0.911 \\
5973.64950 & 3x900  & -23.2 $\pm$ 6.0 &  81.8 $\pm$ 5.8 & 0.112 \\
6312.67329 & 3x900  & -27.2 $\pm$ 4.5 &  90.4 $\pm$ 4.2 & 0.373 \\
6314.84030 & 3x900  &  75.3 $\pm$ 4.4 & -14.5 $\pm$ 4.3 & 0.592 \\
6315.72956 & 3x900  & 104.7 $\pm$ 6.1 & -47.4 $\pm$ 6.5 & 0.682 \\
6622.90358 & 3x900  & 111.3 $\pm$ 4.6 & -54.8 $\pm$ 5.0 & 0.724 \\
6624.82634 & 3x900  &  72.4 $\pm$ 3.8 & -11.1 $\pm$ 4.0 & 0.919 \\
6667.82964 & 3x900  & -52.1 $\pm$ 4.4 & 116.2 $\pm$ 4.6 & 0.265 \\
6998.81780 & 2x900  & 111.4 $\pm$ 4.6 & -51.6 $\pm$ 5.2 & 0.714 \\
6998.92375 & 3x900  & 114.1 $\pm$ 4.7 & -50.8 $\pm$ 5.4 & 0.725 \\
7000.82168 & 3x900  &  73.9 $\pm$ 3.9 & -12.2 $\pm$ 4.0 & 0.916 \\
7285.00910 & 2x900  &  90.0 $\pm$ 4.5 & -34.7 $\pm$ 4.7 & 0.636 \\
7286.00639 & 3x900  & 111.7 $\pm$ 4.5 & -57.1 $\pm$ 4.8 & 0.737 \\
7290.99600 & 3x900  & -55.1 $\pm$ 5.2 & 114.1 $\pm$ 5.0 & 0.241 \\
7359.84854 & 3x900  & -51.5 $\pm$ 5.4 & 110.6 $\pm$ 5.1 & 0.199 \\
7361.88721 & 3x900  & -16.4 $\pm$ 3.7 &  77.0 $\pm$ 4.5 & 0.405 \\
7672.01716 & 3x900  & 109.0 $\pm$ 6.0 & -61.0 $\pm$ 6.1 & 0.747 \\
7672.99925 & 2x900  &  95.9 $\pm$ 5.1 & -41.6 $\pm$ 4.7 & 0.846 \\
7674.98439 & 1x900  &   3.2 $\pm$ 4.1 &  52.1 $\pm$ 4.1 & 0.047 \\
7675.99571 & 3x900  & -41.0 $\pm$ 4.3 &  92.8 $\pm$ 4.5 & 0.149 \\
8859.77310 & 3x900  & 113.8 $\pm$ 5.0 & -51.4 $\pm$ 5.1 & 0.780 \\
8860.79281 & 3x900  &  84.8 $\pm$ 4.5 & -24.9 $\pm$ 4.1 & 0.883 \\

\toprule
\end{tabular}
\end{table}


 \begin{table}
   \caption{Radial velocities near conjunction}
   \label{tab:LogConj}
   \begin{center}
  \begin{tabular}{cccc}
      \hline
HJD      & $t_{\rm exp}$ & RV &  Phase \\
$-2450000$ & (s) & (km~s$^{-1}$) &  ~~~ \\
      \hline

3745.7547 & 3x600  & 28.28 $\pm$ 1.75 & 0.963 \\
3745.8023 & 3x600  & 28.52 $\pm$ 1.81 & 0.968 \\
3745.8499 & 3x600  & 27.70 $\pm$ 1.44 & 0.973 \\
3745.8940 & 3x600  & 26.96 $\pm$ 1.41 & 0.977 \\
4077.6783 & 1x1200 & 28.50 $\pm$ 1.28 & 0.507 \\
4077.7273 & 3x1200 & 29.15 $\pm$ 1.32 & 0.512 \\
4077.7831 & 3x1200 & 28.78 $\pm$ 1.31 & 0.518 \\
4077.8460 & 3x1200 & 28.54 $\pm$ 1.43 & 0.524 \\
4350.0023 & 2x1200 & 29.08 $\pm$ 1.42 & 0.028 \\
5205.6054 & 3x900  & 29.43 $\pm$ 1.39 & 0.494 \\
5205.6427 & 3x900  & 29.24 $\pm$ 1.55 & 0.498 \\
5205.6981 & 4x900  & 30.44 $\pm$ 1.36 & 0.504 \\
5205.7417 & 3x900  & 27.98 $\pm$ 1.43 & 0.508 \\
5205.7776 & 3x900  & 29.59 $\pm$ 1.35 & 0.512 \\
5205.8158 & 3x900  & 28.48 $\pm$ 1.51 & 0.516 \\
5205.8527 & 3x900  & 25.55 $\pm$ 1.41 & 0.519 \\
5205.8992 & 4x900  & 27.72 $\pm$ 1.44 & 0.524 \\
5937.7628 & 1x900  & 28.41 $\pm$ 2.81 & 0.485 \\

  \hline
 \end{tabular}
 \end{center}
\end{table}

\begin{table}
\begin{center}
   \caption{Spectroscopic Orbital Parameters of $\theta^1$\,Ori\,E }
   \label{tab:orbitalpar}
   \begin{tabular}{llll}
      \hline
Parameter & This\,Paper &  Costero\,et\,al. & Herbig\,\&\,Griffin \\
 & & (2008) & (2006) \\
      \noalign{\smallskip}
      \hline
      \hline
      \noalign{\smallskip}

$P_{\rm{orb}}$ (d) & 9.89522(3)  & 9.89520(69) & 9.89456(26) \\
 $\rm{HJD_0}$* & 3281.0455(94) & 3281.039(17)  &  1114.180 \\
$\gamma$ (km~s$^{-1}$)  & 29.7(0.2) & 34.3(0.7) & 30.4(1.0)\\
$K_1$ (km~s$^{-1}$)    & 83.36(0.29) & 84.2(1.2) & 82.4(1.4) \\
$K_2$ (km~s$^{-1}$)    & 84.57(0.29) & 84.6(1.3) & 83.8(1.6) \\
$q = K_1/K_2$ & 0.9856(47) & 0.995(0.018) &  0.9833(0.026) \\

     \noalign{\smallskip}
     \hline
   \end{tabular}
 \end{center}
* HJD--2450000 of the secondary component's inferior conjunction
\end{table}


\subsection{Orbital parameters}
\label{sec:OrbParam}

We determine the orbital parameters of $\theta^1$~Ori~E from the measured radial velocities of both sets of absorption lines. Initially we used the Spectroscopic Binary Solver software \citep{SBS2004} to calculate the orbital parameters of the system, including the eccentricity, and found that it converged to a very small value and is consistent with zero within its uncertainty ($e = 0.002\pm0.002$; see also Section ~\ref{sec:RMEffect} below). In what follows, we assume that the orbit of the system is circular.
Therefore we have fitted sinusoids of the form 
$$V(t) = \gamma + K_{1,2} \sin[2\pi (t - \rm{HJD_0})/P_{\rm orb}]\,,$$
where $\gamma$ is the systemic velocity, $K_{1,2}$ are the semi-amplitude of the components, $\rm{HJD_0}$ the time of inferior conjunction of the secondary and $P_{\rm orb}$ is the orbital period of the binary. We used $\chi^2$ as our goodness-of-fit parameter based on a least-squares algorithm. 
The results of the fit to the individual RV curves are very similar for both components, so we have adopted their averages, when applicable. However, we are able to distinguish the slightly more massive primary star from its companion with 4$\sigma$ certainty. The results are listed in Table~\ref{tab:orbitalpar}, together with the other previously published orbital parameters that, within the errors quoted, are consistent with our more precise results. The only exception is the systemic velocity given by \citet{cea06}, which is significantly higher than that obtained in this paper and the one given by \citet{herfin06}. This difference is due to the fact that \citet{cea06} used the spectra of $\beta$~ Vir ---an F8V spectral type standard--- as a template to obtain the radial velocities of $\theta^1$~Ori~E, without applying zero-point corrections.

\begin{figure*}
\includegraphics[angle=0
,width=1.6\columnwidth]{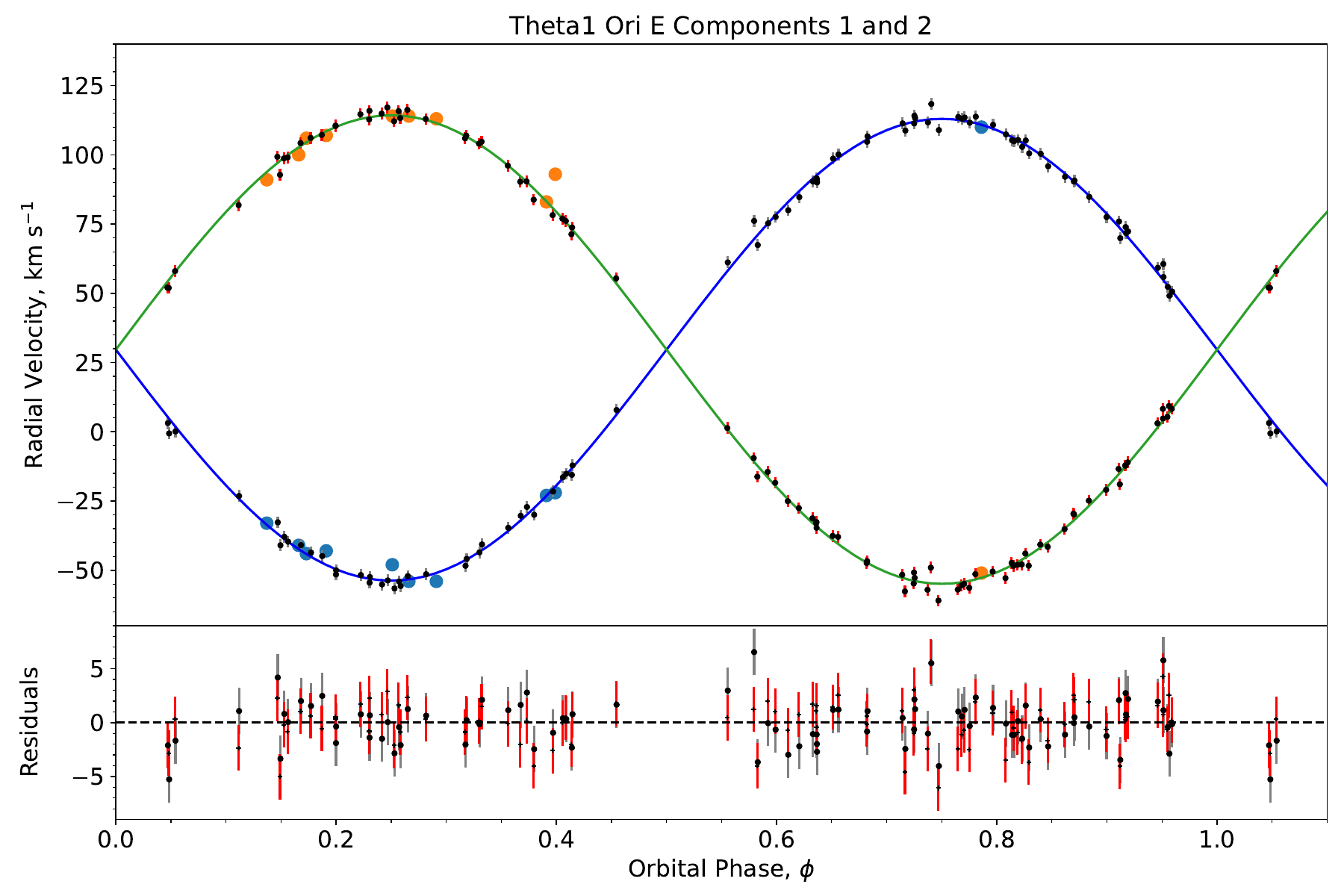}
    \caption{Radial velocity curve of $\theta^1$~Ori~E. Small, black points are our radial velocity points, while large color circles are those obtained by \citet{herfin06}, and are not used in our orbital solution. The blue line is the best fit solution for the primary component; the green one is that for the secondary. Residuals (O-C) are shown in the bottom panel. The red error bars correspond to the primary component, while the gray ones to the secondary.}
\label{fig:radvel}    
\end{figure*}

The resulting velocity curves, folded with the mean orbital period obtained here, are plotted in Figure~\ref{fig:radvel}. For comparison, in addition to our data points, we also plot the data published by \citet{herfin06} (adjusted to our time reference); they all fit very well in our solution except for their very first data point, for which we infer at least a zero-point error of about 5 $km~s^{-1}$. This conjecture arises from the fact that, at any time, the average of the observed radial velocities of the binary components with nearly equal masses must be equal to the systemic velocity of the binary, which is the case for all data points in \citet{herfin06}, except for their first one. 


Our large number of radial velocity data points and the wide time-span during which they were obtained enables us to derive a precise ephemeris of the binary. It must be noted that our $\sim$15-year radial velocity dataset provides a more precise orbital period than that derived from the Spitzer light curves, because the light curves have relatively low photometric precision and show shallow eclipses. Thus, using the values listed in Table~\ref{tab:orbitalpar}, we obtain
\begin{equation}
    HJD = 2453281.0455 \pm 0.0094 + (9.89522 \pm 0.00003)\,E,
\end{equation}
where $HJD$ is the heliocentric Julian Date and $E$ is the cycle given by an integer number. 


\subsection{Binary orbital separation and minimum masses}
\label{sec:MassSep}

To obtain a measurement of the projected orbital separation of $\theta^1$~Ori~E and of the minimum mass of binary components, we use the values of the orbital period, $P_{\rm orb}$, and the semi-amplitudes, $K_{1}$ and $K{_2}$, derived above. 
Our radial velocity measurements of $\theta^1$~Ori~E, have a very good orbital phase coverage and small errors compared to the amplitude of the velocity curves. This allows us to obtain a very precise projected separation of the binary and accurate minimum masses of its stellar components. 
It is important to stress here that we assume that none of the several effects that may alter the observed radial velocity --like mutual irradiation, the hot or cold spots on their photospheres and those caused by accreting circumstellar material-- do not significantly alter our radial velocity measurements and, hence, those of the semiamplitudes $K_1$ and $K_2$, so that they truly represent the orbital movement of the components. Thus, using the values listed in Table~\ref{tab:orbitalpar} and, adopting $e=0$, the separation between the two stars is given by:

\begin{equation}
    a \sin i =\frac { P_{orb} ({ K }_{ 1 }+{ K }_{ 2 }) }{ 2\pi  }=32.84 \pm 0.08 \, R_{\odot}
\end{equation}

\noindent and the minimum masses for the individual components become: 

\begin{equation}
   { M }_1{ \sin }^{ 3 }i=\frac { P_{orb} { K }_{ 2 }({ { K }_{ 1 }+{ K }_{ 2 }) }^{ 2 } }{ 2\pi G }= 2.445 \pm 0.018 \, M_{\odot}\, ;
\end{equation}

\begin{equation}
    { M }_2{ \sin }^{ 3 }i=\frac { P_{orb} { K }_{ 1 }({ { K }_{ 1 }+{ K }_{ 2 }) }^{ 2 } }{ 2\pi G }=2.410 \pm 0.018  \, M_{\odot}\,.
\end{equation}

\smallskip
\smallskip


\subsection{Effective temperature}
\label{sec:Temp}

The effective temperature of stars can be determined by comparing the ratio of pairs of spectral lines observed in their spectra with those obtained for the same lines in suitable synthetic spectra. The ratio of the selected lines should ideally vary strongly with temperature, but not with other parameters such as stellar luminosity. To avoid problems arising from differences in relative chemical abundances, the lines involved in the ratio should also be of the same element, or better yet of the same species. Additionally, to minimize the effects of light contamination arising from the stars themselves (such as veiling due to chromospheric activity or accretion) or from contamination due to external sources, the pair of lines should be as close as possible to each other in wavelength, but separated enough so as not to be badly blended with each other. Of course, each line in the pair should not be blended with other lines that may significantly alter their ratio. Examples of such pairs of lines that have been used as effective temperature diagnostics can be found in \citet{Gray1994} or \cite{Catalano2002}. These authors deal with stars cooler than about 5000~K which are also slow rotators, so the lines selected by these authors are either too weak or strongly mixed with each other in the spectrum of $\theta^1$~Ori~E.

We searched for such pairs of lines in the synthetic spectra described in Section~\ref{sec:RadialVel}. For these purposes, the Catalog of Solar Spectral Lines \citep{moore66} and the NIST Atomic Spectra Data Base \citep{NIST2024} were widely used. The lines in each pair, as mentioned before, were selected to be of the same chemical element but arising from (lower) levels with very different excitation energies. In addition, they were required to have moderate or strong intensities in the relevant temperature range and to be separated by more than 1.5~\AA\,  but by less than about 5~\AA. Our search was limited to the 5200--5700~\AA~wavelength interval in  order to avoid strong nebular and telluric lines. We found only two such pairs, namely:
\begin{itemize}
\item Fe\,I\,5447/Fe\,I\,5445, formed mainly by the Fe\,I $\lambda\lambda$\,5446.49, 5446.92 and 5445.04 \AA\, lines, with lower excitation levels of 4.4, 1.0 and 4.4 $eV$, respectively; and
\item Fe\,I\,5404/Fe\,I\,5406, consisting of Fe\,I $\lambda\lambda$\,5403.82, 5404.15, 5405.35 and 5405.78, whose lower excitation levels are 4.1, 4.4, 4.4 and 1.0 $eV$, respectively.
\end{itemize}

\begin{figure}

\includegraphics[width=12cm]{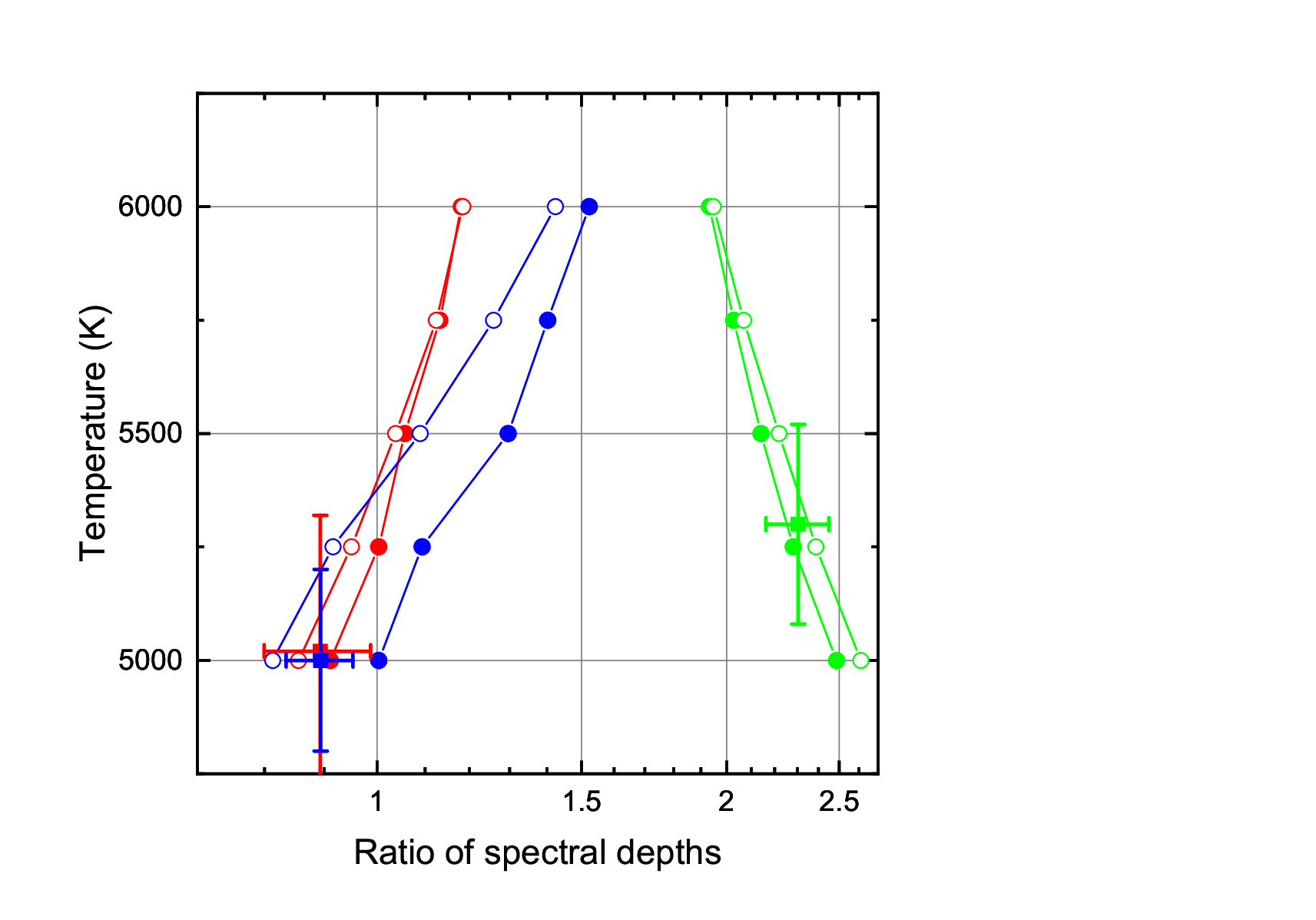}
    \caption{Temperature versus ratio of spectral depths. In green FeI 5447/FeI 5445, in red FeI 5404/FeI 5406, in blue FeII 5317/FeI 5324. Filled dots are for $log\ g=3.0$ and open dots are for $log\ g=3.5$ in the synthetic spectra. The average value for each ratio as measured in the composite spectra of $\theta^1$~Ori~E near conjunction (Table \ref{tab:LogConj} is plotted as a square with horizontal and vertical error bars at an interpolated $log\ g=3.3$, and at estimated Temperature values of $5300\ K$ (green), $5020\ K$ (red) and $5000\ K$ (blue), respectively. From this analysis, we adopt $T_{eff}=5150 \pm 200 \,K$ for the average temperature of both nearly identical components of the binary.}
    \label{fig:TvsDEPTHLOG}
\end{figure}

In Figure \ref{fig:TvsDEPTHLOG}, we present a graph of the ratio of the depths of these lines versus the effective temperature, as measured in the synthetic spectra, with $log\ g=3.0$ (filled dots) and $log\ g=3.5$ (open dots). We have used the depth (and not the equivalent width) of the Gaussian fitted to the lines by the {\sc IRAF} {\it splot} task because the results are less noisy, probably as a consequence of blending with weaker lines that alter the depth of the lines less than their equivalent width. In doing so, we used the {\it deblend} option of {\it splot}.
In the figure, the ratio Fe\,I 5447/Fe\,I 5445 is represented in green, and that of Fe\,I 5404/Fe\,I 5406 in red. 
The same ratios were also measured in the spectra of $\theta^1$~Ori~E near conjunction, listed in Table \ref{tab:LogConj}, except for the first and last ones because the line ratios obtained from them are often very different from the rest, probably due to the relatively higher light contamination in those two spectra. The average of these ratios are, respectively, $2.305 \pm 0.144$ (green) and $0.893 \pm 0.094$ (red), the errors being the standard deviation from the mean. Interpolating for the value of $log\ g=3.3$, we estimate the respective temperatures $5300 \pm 220\ K$ (green) and $5020 \pm 300\ K$ (red), represented in the graph as colored squares with vertical and horizontal error bars. Their average, $5160 \pm 260\,K$, is in excellent agreement with the estimate of \citet{herfin06}, $T \sim 5100\,K$, from the spectral type of this star. 

The ratios of lines arising from the same element, but at different ionization stages, may also be used to estimate stellar temperatures, although they are usually also very sensitive to the surface gravity of the stars. Furthermore, due to chromospheric activity (known to be present in both components of $\theta^1$~Ori~E, inferred from the Ca\,II\,K emission in their spectra), the Fe\,II lines may be partially filled in and their ratio with the Fe, I lines may produce lower temperatures than those derived from the line ratios of neutral species of the same element.
One such pair of hybrid ionization state spectral lines is Fe\,II\,5317/Fe\,I\,5324, integrated mainly by Fe\,II $\lambda\lambda$\,5316.62 and 5316.78, and Fe\,I $\lambda$\,5324.19, and is also shown in Figure \ref{fig:TvsDEPTHLOG} in blue. Its strong dependence on $log\ g$, mainly at lower temperatures, is evident. An additional caveat is the somewhat large separation in wavelength between the line pair.
Not used to compute the temperature of the binary, we found that their average ratio in spectra near conjunction, $0.894 \pm 0.059$,  yields a temperature of $5000 \pm 200\ K$, consistent with that derived above. 

We hereafter adopt $T_{\rm eff}=5150 \pm 200 \,K$ for the average temperature of both nearly identical components of the binary.

\subsection{Rotational velocity}
\label{sec:RotatVel}

We have measured the projected rotational velocities of the components of $\theta^1$~Ori~E by comparing a fixed width near quadrature with broadened versions of the standard stars $\beta$~Vir (F8\,IV) and $\xi$~Boo~A (G8\,V) --both very slow rotators-- following the procedure described in detail by \citet{eal2008}. In this way, we find $v\,sin i = 32 \pm 3$~km~s$^{-1}$, similar to the value estimated by \citet{herfin06} of $35 \pm 5~$km~s$^{-1}$.

The latter authors consider this value to be consistent with the binary components being in rotational synchronization. Indeed, using the orbital period obtained here and the \citet{mor12} results for this eclipsing binary (namely $R_1+R_2 = 12.5  ~\pm \, 0.6\, R_\odot$, for the sum of the radii of the two stars, and $i = \, 73.7\, \pm \, 0.9 ^{\circ}$ for the projected orbital inclination), we obtain $v\,sin i~ = 30.7 \,\pm \, 2.1~km~s^{-1}$ if the rotational period and the orbital one are equal. This value is in excellent agreement with the one determined here. In the above calculation, we have assumed that the radii of the two stars are equal and that, as justified in Section \ref{sec:RMEffect}, their orbit and equatorial planes are co-planar.

We hence conclude that the binary components are in rotational synchronization.

\subsection{Search for the (small) Rossiter-McLaughlin (RM) Effect and the orbital eccentricity}
\label{sec:RMEffect}

The spectra listed in Table~\ref{tab:LogConj} were obtained in order to possibly detect the RM effect \citep{Rossiter1924, McLaughlin1924} in $\theta^1$~Ori~E, produced during an as yet undetected mutual eclipse in what had then been found to be a rather close spectroscopic binary with giant stellar components. Through the work by \citet{mor12}, we now know that the effect cannot be revealed by our observations, as shown in what follows.

\begin{figure}
	
	\includegraphics[width=1.0\columnwidth]{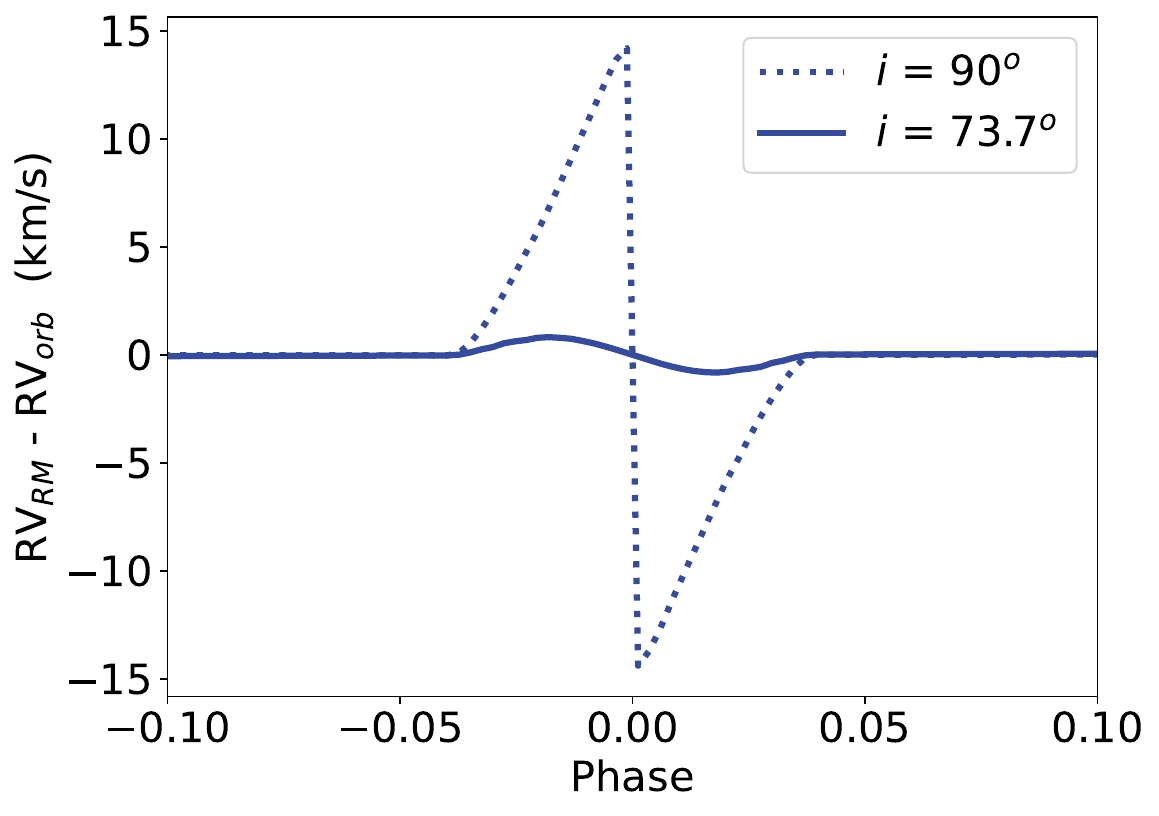}
    \caption{Predicted amplitude of the Rossiter-McLaughlin effect during the primary eclipse of $\theta^1$~Ori~E. We show the predicted deviation of the radial velocity curve from the Keplerian orbit during the primary transit with the best-fit radii and inclination from \citet{mor12} (solid line) and the upper limit of the deviation (i= 90$^\circ$; dotted line). The radial velocities presented in this paper at conjunction are not sufficiently precise to measure the RM effect. }

    \label{fig:th1OriE_rm}
\end{figure}

Using the eclipsing binary (EB) code {\sc phoebe} \citep{Prsa2005}, we modeled the RM effect to assess the magnitude of the deviation of the radial velocity from the pure Keplerian orbit during the primary eclipse. Because the orbit is circular and the timescale for circularization of the orbit is longer than both the timescale for alignment and the timescale for rotational synchronization \citep[and references therein]{GomezMaqueoChew2012}, we assume that both stellar components are aligned and rotating synchronously with the orbital period (also see Section~\ref{sec:RotatVel}), the latter being robustly measured from the well-sampled radial velocity curves. We adopted the best values for the inclination and sum of the radii derived from the Spitzer light curves in \citet{mor12} for our RM model (see Table~\ref{table:properties}), and assume, given their almost identical masses, that the radii are equal (R$_1$ $\approx$ R$_2$). Given these constraints from the Spitzer light curves, the modeled RM effect has a peak-to-peak amplitude of 1.6\kms. It must be noted \citep[as shown in Figure~9 of ][]{mor12} that the sum of the radii and the inclination are highly degenerate, given the relatively low sampling and the high dispersion of the eclipse light curves. Thus, we also considered an inclination of 90$^\circ$ (and R$_1$+R$_2$=~8~R$_{\odot}$) in order to preserve the observed eclipse duration) to set an upper limit for the amplitude of the RM effect, which is modeled to have a peak-to-peak amplitude of 28.5$\,km~s^{-1}$. The anomaly of the RV curve due to the RM effect is shown in Fig.~\ref{fig:th1OriE_rm} for both the best fit model of \citet{mor12} (solid line) and the upper limit case of $90^\circ$ (dotted line).  

As a check on whether the circular orbit assumption is correct, we fitted the light curves and radial velocities with {\sc Phoebe} and let the eccentricity, angle of periastron ($\omega$) and the phase shift be free parameters. We derive an eccentricity of 0.0033 $\pm$ 0.0029, which has a probability of $\sim$ 52\% of being spurious, as defined by \citet{Lucy1971}. The most likely scenario given the available data is that the orbit is circular. 

\section{Discussion}
\label{sec:conclusions}

The main purpose of this paper has been to obtain precise radial velocity curves of the components of $\theta^1$~Ori~E and, hence, the orbital and physical parameters of the binary. The results are addressed in Section~\ref{sec:ResDisc}. The more than fifteen years that our observation span and the applied nebular zero-point corrections allowed us to improve the orbital period and obtain accurate values of the velocity curve semi-amplitudes, the systemic velocity, and the physical scale of the orbit. These basic results are shown in Table~\ref{tab:orbitalpar}, where we also show, for comparison, the results of \citet{herfin06} and \citet{cos08}. The improvement in these orbital parameters, in turn, enables us to make a more precise determination of the physical parameters of the system, derived from the previously published light curve \citep{mor12} and, together with PMS evolutionary models, to estimate the age of this binary star. In addition, the more precise derivation of the systemic velocity and the existence of modern, very precise values of stellar proper motions compel us to reconsider the alleged ejection of $\theta^1$~Ori~E from the Orion Trapezium, which would set this binary beyond the reach of the gravitational well of the stellar group. All this is explained in what follows.

\subsection{Updating light curve dependent properties}
Following the analysis presented in \S\ref{sec:RMEffect} with {\sc Phoebe}, we use the eclipsing binary model to update the light-curve-dependent properties and better characterize their uncertainties given the refined orbital parameters measured from the well-populated radial velocity curves presented in this paper. We present the derived physical properties resulting from this analysis in the bottom part of Table~\ref{table:properties}.

\begin{table*}
\caption{Physical properties of the $\theta^1$~Ori~E eclipsing stars}             
\centering          
\begin{tabular}{l c c c }     
\hline\hline       
\multicolumn{2}{c}{Parameter} & Value & Units\\ 
\hline               
\multicolumn{4}{c}{Derived from line ratios (Section~\ref{sec:Temp})}\\
Effective temperature & \teffp\ or \teffs & 5150 $\pm$ 200 & K\\ 
\multicolumn{4}{c}{Adopted from \citet{mor12}}\\
Inclination & $i$ (fixed) &    73.7 $\pm$ 0.9  & \degree  \\
Sum of the radii & \rprim + \rsec\ (fixed) & 12.5 $\pm$ 0.6  & \rsun\\ 
\hline
\multicolumn{4}{c}{Updated light-curve dependent properties (Section\ref{sec:conclusions})}\\
Semimajor axis & $a$ & 34.22  $\pm$  0.18  & \rsun\\ 
 & $a$ &  0.1589  $\pm$  0.0008 & au \\  
Primary mass & \mprim\  &  2.755  $\pm$  0.043  & \msun \\
Secondary mass & \msec &  2.720  $\pm$  0.043 & \msun \\
Primary radius & \rprim\  & 6.26  $\pm$  0.31  & \rsun \\ 
Secondary radius & \rsec &  6.25  $\pm$  0.30 & \rsun \\ 
Primary surface gravity & \loggp  &    3.29  $\pm$  0.04 & dex (cgs) \\ %
Secondary surface gravity & \loggs &   3.28  $\pm$  0.04 & dex (cgs) \\ %
Temperature ratio & \teffs/\teffp &  1.000  $\pm$  0.040 & \\
Primary luminosity  & \ensuremath{L_{\mathrm {1}}}  &   24.8  $\pm$  4.6  & L$_{\odot}$ \\  
Secondary luminosity  & \ensuremath{L_{\mathrm {2}}} &   24.7  $\pm$  4.5 & L$_{\odot}$\\  
\hline       
\end{tabular}
\label{table:properties}
\end{table*} 

    \begin{figure}
	
	\includegraphics[width=9cm,height=9cm]{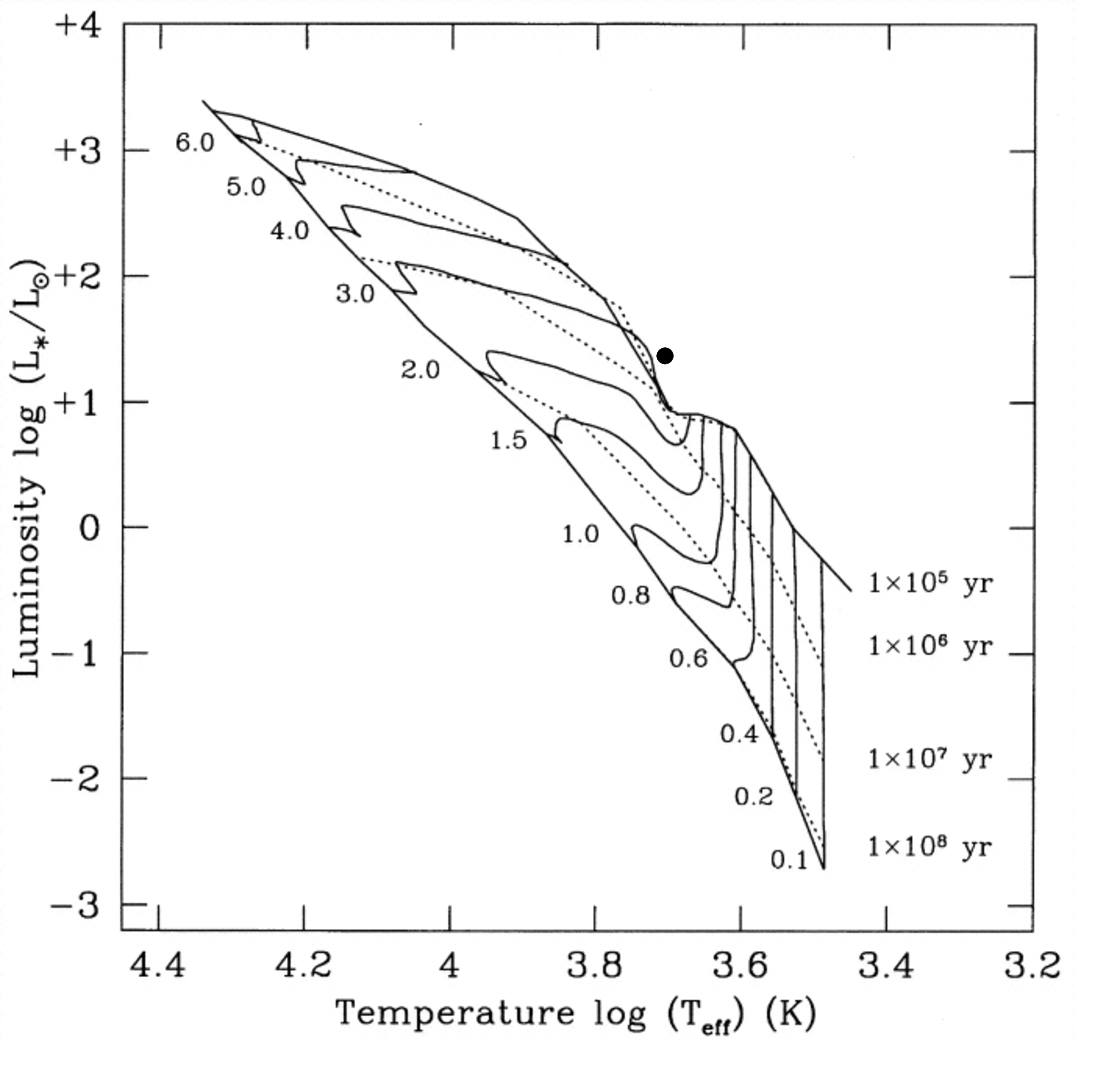}
    \caption{Pre-main sequence evolutionary models from \citet{Palla1999}: We have adapted their Fig.~1 by placing a solid dot at our measurement of $\theta^1~Ori~E$ ($L=24.8\pm 4.6\ L_{\odot}$ and $T_{eff} = 5150 \,  \pm \, 200 \ K$). The uncertainty in the luminosity is roughly the size of the point, while the uncertainty in the temperature is much smaller than the symbol. The dotted lines are isochrones that span from 0.5 to 100 Myr. The solid black lines describe the evolution of stars of different masses, where the mass (in \msun) is given where the PMS evolutionary track reaches the Zero-Age Main Sequence. Importantly, the evolutionary tracks start at the `birth-line', that was empirically defined for the Orion Nebula Cluster in \citet{Palla1999}, and that describes the moment in the star's PMS stage at which the envelope becomes transparent and thus the star emerges from the molecular cloud.}
    \label{fig:hrd}
    \end{figure}


\subsection{The age of \texorpdfstring{$\theta^1~\mathrm{Ori}~E$}{theta1 Ori E} and the Orion Trapezium}
\label{sec:evolution}

Our determinations of $log\ \teff$ and $log\ L/L_\odot$ of the nearly identical components of $\theta^1~Ori~E$ allow us to precisely locate the star in the HR diagram, as shown in Fig.~\ref{fig:hrd}. There, the evolutionary tracks of the PMS stars, calculated by \citet{Palla1999}, are included. We have used these evolution models because the birth-line adopted by these authors (when the stars emerge from the molecular cloud in which they form) has been empirically calibrated for the Orion Nebula Cluster. 
In the HR diagram, the $\theta^1~\mathrm{Ori}~E$ measurement is slightly above both the $10^5$ year isochrone and the birth-line. This implies that the object is somewhat younger than $10^5 yr$ and that its envelope was prematurely dissipated. We speculate that such a seemingly `premature' appearance might have been caused by the strong radiation of the nearby, much brighter and hotter components of the Orion Trapezium, which would evolve faster and cause the early loss of the envelope around the lower-mass close neighbors, and hence their consequent precocious emergence from the molecular cloud.

As for the age of $\theta^1~Ori~E$, while undoubtedly very young, a careful analysis of the influence that certain parameters may have on the age determination of such very young stars --like the stellar rotation and the chemical composition-- must be done and is beyond the scope of this paper. However, there are several other clues to the extreme youth of the Orion Nebula and Trapezium. Examples of this are:
\begin{enumerate}
    \item \citet{PflammKroupa2006} concluded that the Orion trapezium should have dissipated by now; 
    \item \citet{O'Delletal2009} estimate the age of the ‘Huygens region’ of the Orion Nebula to be 15 000 yr; 
    \item the dynamical age for Component B of the Trapezium -- a mini-cluster now known to consist of at least 6 stellar components -- was found to be about 30 000 yr by \citet{Allenetal2015}; 
    \item \citet{Allenetal2017} found that the dynamical lifetime of the Trapezium as a whole is between 10 000 and 30 000 yr,
    \item the Trapezium Component A (V1016 Ori), an eclipsing and spectroscopic binary, has been found by \citet{Valle2011} and \citet{Costero2019} to have a highly inclined orbit (obliquity with respect to the equatorial plane of the primary star), indicative of primordial misalignment (that shortly disappears) or a very recent dynamical interaction.
\end{enumerate}

In conclusion, the extreme youth of $\theta^1~Ori~E$ --implied by its position in the HR diagram and compared to the evolutionary tracks of the PMS -- is probably real and worth a more careful and specialized analysis.

\subsection{Separation from Component A}
\label{sec:Separation}

 As already mentioned in Section~\ref{sec:intro}, \citet{apw74} and \citet{oli13} concluded that the Orion Trapezium components E and A are separating from each other at a rate slightly larger than the escape velocity of the Orion Trapezium. They based this conclusion mainly on historical measurements of the relative position of the two stars and, in the case of the latter authors, diffracto-astrometric measurements of the Trapezium stars obtained from WFP-C2 HST archival images. This conclusion was supported by the systemic radial velocity obtained by \citet{cos08} for $\theta^1~Ori~E$ (see Table~\ref{tab:orbitalpar}), which is approximately 6~km~s$^{-1}$ higher than that of Component A, which according to \citet{Vitrichenkoetal1998} and \citet{SticklandLloyd2000} is $28\pm1$ km~s$^{-1}$. However, the systemic velocity of Component E obtained by us in this work and the one published by \citet{herfin06} (see Table~\ref{tab:orbitalpar}), are only about 2 km~s$^{-1}$ larger than that of Component A, also an eclipsing binary, putting into doubt the escape of E from the Trapezium. 

It is therefore appropriate to consider the separation rate of Component E from Component A in light of modern astrometric results, from which proper stellar motions can be obtained with a precision better than one $mas/yr$. We refer to the {\it third Gaia data release} 
\citep[\rm{DR3};][]{Gaia2016,GaiaDR32023}
and the Very Large Base Array (\rm{VLBA)} series of observations published by \citet{Dzibetal2021}. 

The proper motions of Component E derived from \rm{DR3} are $\mu_{\alpha*}(E)=1.345\pm0.022\ mas/yr$ and $\mu_{\delta}(E) = 1.186\pm 0.021\  mas/yr$, while those of Component A are $\mu_{\alpha*}(A)=1.355\pm\ 0.058\ mas/yr$ and $\mu_{\delta}(A)=0.250\pm0.048\ mas/yr$, in right ascension and declination, respectively. Their respective differences are the components of transverse separation of E relative to A: $\Delta\mu_{\alpha*}(E-A)=-0.010\pm 0.062\ mas/yr$ and $\Delta \mu_{\delta}(E-A) = 0.936\pm 0.052\ mas/yr$, or a total transverse angular separation rate of $0.936\pm 0.081\ mas/yr$. Adopting the distance to the ONC derived from \citet{Maizetal2022} from \rm{Early DR3} data, $390\pm2 pc$ (essentially equal to that determined by \citet{Kounkeletal2017}, $388\pm5\ pc$, based on \rm{VLBA} observations), a transverse velocity of $1.74\pm\ 0.15\ km~s^{-1}$ is obtained.

This result, derived from the data obtained by Gaia in a crowded and bright nebular region, may be affected by systematic errors \citep[e.g.,][]{Weietal2024}. Hence, it is appropriate to compare them with those obtained from the proper motions derived from a completely independent source, with comparable precision. This is the case of those given by \citet{Dzibetal2021}, derived from \rm{VLBI} astrometry. 
Both $\theta^1~Ori~E$ and $\theta^1~Ori~A$ are radio sources. The radio source of  Component A, labeled $A_{2}$ in \citet{Dzibetal2021}, is a weak tertiary component discovered by \citet{Petretal1998}, located 0.2 arcsec north of the bright eclipsing and spectroscopic binary (V1016 Ori; frequently labeled $A_{1,3}$).
The relative position of this tertiary star (generally labeled $A_{2}$) with respect to the primary component has been measured over a time span of about 20 years using various interferometric methods; the results were compiled by \citet{Karletal2018} and plotted in their Figure 2 as a short segment of its hypothetical orbit around $A_{1,3}$. A simple linear fit to these data points yields an average relative motion rate of $A_2$ with respect to $A_{1,3}$, which we estimate to be $\Delta\alpha_*(A_2)=3.8\pm0.1\ mas/yr$ and $\Delta\delta(A_2)=-2.5\pm0.1\ mas/yr$.
The proper motion of $A_2$ given by \citet{Dzibetal2021}, $\mu_{\alpha*}(A_2) = 4.87\pm0.07 \ mas/yr$ and $\mu_{\delta}(A_2)=-2.56\pm0.12\ mas/yr$, must be corrected for this orbital yearly movement of the tertiary (the radio source) to derive the proper motion of the primary $A_{1,3}$: $\mu_{\alpha*}(A)=1.07 \pm0.12\ mas/yr$ and $\mu_{\delta}(A)=-0.06\pm0.16\ mas/yr$.
This proper motion can now be appropriately compared with that given for Component E by \citet{Dzibetal2021}, $\mu_{\alpha*}(E)=1.31\pm0.05\ mas/yr$  and $\mu_{\delta}(E)=1.11\pm0.14\ mas/yr$, to obtain the separation rate of E with respect to A: $\Delta\mu_{\alpha*}(E-A)=0.24\pm0.13\ mas/yr$ and $\Delta\mu_{\delta}(E-A)=1.17\pm 0.21\ mas/yr$, and a total transverse separation rate of $1.19\pm0.25\ mas/yr =2.2\pm0.5\ km~s^{-1}$ at 390 pc, in good agreement with the results derived above from Gaia \rm{DR3} data. 

The separation rates derived here --the one obtained from Gaia \rm{DR3} ($1.74\pm\ 0.15\ km~s^{-1}$) and that from VLBI data ($2.2\pm0.5\ km~s^{-1}$) -- are smaller than the dispersion of the transverse velocity measured in ONC radio sources by \citet{Dzibetal2021} ($2.87\pm0.24\ km~s^{-1}$), and less than that obtained from optical and infrared observations of many ONC members by \citet{Theisenetal2022} ($2.61\pm0.18\ km~s^{-1}$) or by \citet{Weietal2024} $(2.67\pm0.14\ km\ s^{-1})$. The space velocity of $\theta^1~Ori~E$ with respect to $\theta^1~Ori~A$ --- as derived from the proper motions given in Gaia \rm{DR3} and from the relative systemic velocity of both binaries ($1.7\pm1.1\ km~s^{-1}$) --- is certainly less than the escape velocity of Component E from the Orion Trapezium, estimated to be approximately $6\ km~s^{-1}$ \citep[]{apw74, maiz21}. 
We conclude that Component E probably is not escaping from the Orion Trapezium, in agreement with \citet{maiz21} who, using \rm{Early DR3} observations, find that Component F is moving away from Component C somewhat faster than the escape velocity, whereas this behavior is not observed in any of the other members of the Orion Trapezium.
 

\section{Summary}
\label{sec:sum}

We confirm that $\theta^1$~Ori~E is a PMS, double-lined spectroscopic and eclipsing binary, with almost identical stellar components in synchronous rotation with its orbital period. We find that their masses are significantly different from each other (q = 0.9856 $pm$0.0047). The binary components are probably the most massive stars known with masses determined with precision better than 2 percent both being PMS stars. 

Using the previously published light curve and orbital inclination, together with our results from the velocity curves and the estimate of effective temperature, we have placed the star in the H-R diagram together with theoretical PMS evolutionary models; a very young age, probably younger than $10^5 yr$, is found for the binary, in agreement with additional indicators of the extreme youth of the Orion Trapezium. 

A better determination of the effective temperature of the components and a more precise and better sampled light curve are needed to further test evolutionary models of intermediate-mass PMS stars.


\section*{Acknowledgments}
{ The authors thank the anonymous referee for the careful review and kind comments.} 
Based on observations obtained at the Observatorio Astron\'omico Nacional at San Pedro M\'artir, Baja California, M\'exico, operated by the Instituto de Astronom\'{\i}a, Universidad Nacional Aut\'onoma de M\'exico.
We thank Julio Clemente and Ivan Mora Zamora for their help. 
YGMC has been partially supported by UNAM-PAPIIT-IG101224. JE is indebted for the support of DGAPA-UNAM through project PAPIIT IN113723. ARM also thanks  DGAPA at UNAM for financial support under projects PAPIIT IN103813, IN102517 and IN102617.

\section*{Data availability}

The data underlying this article can be shared on request with the corresponding author.


\bibliographystyle{mnras}
\bibliography{bibliography.bib} 

@ARTICLE{mor12,
       author = {{Morales-Calder{\'o}n}, M. and {Stauffer}, J.~R. and {Stassun}, K.~G. and {Vrba}, F.~J. and {Prato}, L. and {Hillenbrand}, L.~A. and {Terebey}, S. and {Covey}, K.~R. and {Rebull}, L.~M. and {Terndrup}, D.~M. and {Gutermuth}, R. and {Song}, I. and {Plavchan}, P. and {Carpenter}, J.~M. and {Marchis}, F. and {Garc{\'\i}a}, E.~V. and {Margheim}, S. and {Luhman}, K.~L. and {Angione}, J. and {Irwin}, J.~M.},
        title = "{YSOVAR: Six Pre-main-sequence Eclipsing Binaries in the Orion Nebula Cluster}",
      journal = {\apj},
         year = 2012,
        month = jul,
       volume = {753},
       number = {2},
          eid = {149},
        pages = {149},
          doi = {10.1088/0004-637X/753/2/149},
archivePrefix = {arXiv},
       eprint = {1206.6350},
 primaryClass = {astro-ph.GA},
       adsurl = {https://ui.adsabs.harvard.edu/abs/2012ApJ...753..149M}
}

@ARTICLE{Catalano2002,
       author = {{Catalano}, S. and {Biazzo}, K. and {Frasca}, A. and {Marilli}, E.},
        title = "{Measuring starspot temperature from line depth ratios. I. The method}",
      journal = {\aap},
         year = 2002,
        month = nov,
       volume = {394},
        pages = {1009-1021},
          doi = {10.1051/0004-6361:20021223},
       adsurl = {https://ui.adsabs.harvard.edu/abs/2002A&A...394.1009C}
}

@ARTICLE{esal04,
       author = {{Esteban}, C. and {Peimbert}, M. and {Garc{\'\i}a-Rojas}, J. and {Ruiz}, M.~T. and {Peimbert}, A. and {Rodr{\'\i}guez}, M.},
        title = "{A reappraisal of the chemical composition of the Orion nebula based on Very Large Telescope echelle spectrophotometry}",
      journal = {\mnras},
         year = 2004,
        month = nov,
       volume = {355},
       number = {1},
        pages = {229-247},
          doi = {10.1111/j.1365-2966.2004.08313.x},
archivePrefix = {arXiv},
       eprint = {astro-ph/0408249},
 primaryClass = {astro-ph},
       adsurl = {https://ui.adsabs.harvard.edu/abs/2004MNRAS.355..229E}
}

@ARTICLE{herfin06,
       author = {{Herbig}, G.~H. and {Griffin}, R.~F.},
        title = "{{\ensuremath{\theta}}$^{1}$ Orionis E as a Spectroscopic Binary}",
      journal = {\aj},
         year = 2006,
        month = nov,
       volume = {132},
       number = {5},
        pages = {1763-1767},
          doi = {10.1086/507769},
       adsurl = {https://ui.adsabs.harvard.edu/abs/2006AJ....132.1763H}
}

@ARTICLE{cea06,
       author = {{Costero}, R. and {Echevarria}, J. and {Richer}, M. and {Poveda}, A.},
        title = "{theta\^1 Orionis E}",
      journal = {\iaucirc},
         year = 2006,
        month = feb,
       volume = {8669},
        pages = {2},
       adsurl = {https://ui.adsabs.harvard.edu/abs/2006IAUC.8669....2C}
}

@ARTICLE{par54,
       author = {{Parenago}, Pavel Petrovich},
        title = "{Issledovaniia zvezd v oblasti tumannosti Oriona}",
      journal = {Trudy Gosudarstvennogo Astronomicheskogo Instituta},
         year = 1954,
        month = jan,
       volume = {25},
        pages = {3-543},
       adsurl = {https://ui.adsabs.harvard.edu/abs/1954TrSht..25....1P}
}

@ARTICLE{her50,
       author = {{Herbig}, George H.},
        title = "{Spectra of Variable Stars in the Orion Nebula.}",
      journal = {\apj},
         year = 1950,
        month = jan,
       volume = {111},
        pages = {15},
          doi = {10.1086/145233},
       adsurl = {https://ui.adsabs.harvard.edu/abs/1950ApJ...111...15H}
}

@ARTICLE{apw74,
       author = {{Allen}, C. and {Poveda}, A. and {Worley}, C.~E.},
        title = "{The kinematics of trapezium systems.}",
      journal = {\rmxaa},
         year = 1974,
        month = nov,
       volume = {1},
        pages = {101-118},
       adsurl = {https://ui.adsabs.harvard.edu/abs/1974RMxAA...1..101A}
}

@ARTICLE{walker77,
       author = {{Walker}, M.~F.},
        title = "{Observations of Primary Minimum of theta1 Ori A}",
      journal = {Information Bulletin on Variable Stars},
         year = 1977,
        month = feb,
       volume = {1238},
        pages = {1},
       adsurl = {https://ui.adsabs.harvard.edu/abs/1977IBVS.1238....1W}
}

@article{FaG78,
	doi = {10.1086/130427},
	url = {https://doi.org/10.1086/130427},
	year = 1978,
	month = {dec},
	publisher = {{IOP} Publishing},
	volume = {90},
	pages = {762},
	author = {W. A. Feibelman and T. R. Gull},
	title = {Photographic observations of Theta-1 Orionis},
	journal = {Publications of the Astronomical Society of the Pacific}
}

@BOOK{kak82,
       author = {{Kukarkin}, B.~V. and {Kholopov}, P.~N.},
        title = "{New catalogue of suspected variable stars}",
         year = 1982,
         publisher={Moscow: Publication Office ``Nauka"},
       adsurl = {https://ui.adsabs.harvard.edu/abs/1982ncsv.book.....K}
}

@ARTICLE{wolf94,
       author = {{Wolf}, George W.},
        title = "{CCD Photometry of BM Orionis and Other Trapezium Stars}",
      journal = {Experimental Astronomy},
         year = 1994,
        month = mar,
       volume = {5},
       number = {1-2},
        pages = {61-66},
          doi = {10.1007/BF01583811},
       adsurl = {https://ui.adsabs.harvard.edu/abs/1994ExA.....5...61W}
}

@ARTICLE{oli13,
       author = {{Olivares}, J. and {S{\'a}nchez}, L.~J. and {Ruelas-Mayorga}, A. and {Allen}, C. and {Costero}, R. and {Poveda}, A.},
        title = "{Kinematics of the Orion Trapezium Based on Diffracto-Astrometry and Historical Data}",
      journal = {\aj},
         year = 2013,
        month = nov,
       volume = {146},
       number = {5},
          eid = {106},
        pages = {106},
          doi = {10.1088/0004-6256/146/5/106},
archivePrefix = {arXiv},
       eprint = {1310.0769},
 primaryClass = {astro-ph.SR},
       adsurl = {https://ui.adsabs.harvard.edu/abs/2013AJ....146..106O}
}

@BOOK{webb1859,
       author = {{Webb}, T.~W.},
        title = "{Celestial Objects for Common Telescopes}",
         year = 1859,
         publisher = {London: Longman, Green, Longman, and Roberts.},
       adsurl = {https://ui.adsabs.harvard.edu/abs/1859cect.book.....W}
}

@ARTICLE{gled1880,
       author = {{Gledhill}, Joseph.},
        title = "{Correspondence - The Trapezium in Orion.}",
      journal = {Astronomical register},
         year = 1880,   
        month = jan,
       volume = {18},
        pages = {64-66},
       adsurl = {https://ui.adsabs.harvard.edu/abs/1880AReg...18...64G}
}

@ARTICLE{eal2008,
       author = {{Echevarr{\'\i}a}, J. and {Smith}, Robert Connon and {Costero}, R. and {Zharikov}, S. and {Michel}, R.},
        title = "{High-dispersion absorption-line spectroscopy of AE Aqr}",
      journal = {\mnras},
         year = 2008,
        month = jul,
       volume = {387},
       number = {4},
        pages = {1563-1574},
          doi = {10.1111/j.1365-2966.2008.13248.x},
archivePrefix = {arXiv},
       eprint = {0804.0291},
 primaryClass = {astro-ph},
       adsurl = {https://ui.adsabs.harvard.edu/abs/2008MNRAS.387.1563E}
}

@ARTICLE{kea82,
       author = {{Ku}, W.~H. -M. and {Righini-Cohen}, G. and {Simon}, M.},
        title = "{High-Resolution X-ray Observations of the Orion Nebula}",
      journal = {Science},
         year = 1982,
        month = jan,
       volume = {215},
       number = {4528},
        pages = {61-64},
          doi = {10.1126/science.215.4528.61},
       adsurl = {https://ui.adsabs.harvard.edu/abs/1982Sci...215...61K}
}

@ARTICLE{gea05,
       author = {{Getman}, K.~V. and {Flaccomio}, E. and {Broos}, P.~S. and {Grosso}, N. and {Tsujimoto}, M. and {Townsley}, L. and {Garmire}, G.~P. and {Kastner}, J. and {Li}, J. and {Harnden}, F.~R., Jr. and {Wolk}, S. and {Murray}, S.~S. and {Lada}, C.~J. and {Muench}, A.~A. and {McCaughrean}, M.~J. and {Meeus}, G. and {Damiani}, F. and {Micela}, G. and {Sciortino}, S. and {Bally}, J. and {Hillenbrand}, L.~A. and {Herbst}, W. and {Preibisch}, T. and {Feigelson}, E.~D.},
        title = "{Chandra Orion Ultradeep Project: Observations and Source Lists}",
      journal = {\apjs},
         year = 2005,
        month = oct,
       volume = {160},
       number = {2},
        pages = {319-352},
          doi = {10.1086/432092},
archivePrefix = {arXiv},
       eprint = {astro-ph/0410136},
 primaryClass = {astro-ph},
       adsurl = {https://ui.adsabs.harvard.edu/abs/2005ApJS..160..319G}
}

@ARTICLE{zea04,
       author = {{Zapata}, Luis A. and {Rodr{\'\i}guez}, Luis F. and {Kurtz}, Stanley E. and {O'Dell}, C.~R.},
        title = "{Compact Radio Sources in Orion: New Detections, Time Variability, and Objects in OMC-1S}",
      journal = {\aj},
         year = 2004,
        month = apr,
       volume = {127},
       number = {4},
        pages = {2252-2261},
          doi = {10.1086/382715},
archivePrefix = {arXiv},
       eprint = {astro-ph/0403350},
 primaryClass = {astro-ph},
       adsurl = {https://ui.adsabs.harvard.edu/abs/2004AJ....127.2252Z}
}

@ARTICLE{fea93,
       author = {{Felli}, M. and {Taylor}, G.~B. and {Catarzi}, M. and {Churchwell}, E. and {Kurtz}, S.},
        title = "{The Orion radio zoo revisited: source variability.}",
      journal = {\aaps},
         year = 1993,
        month = oct,
       volume = {101},
        pages = {127-151},
       adsurl = {https://ui.adsabs.harvard.edu/abs/1993A&AS..101..127F}
}

@ARTICLE{cas88,
       author = {{Castaneda}, Hector O.},
        title = "{The Velocity Structure and Turbulence at the Center of the Orion Nebula}",
      journal = {\apjs},
         year = 1988,
        month = may,
       volume = {67},
        pages = {93},
          doi = {10.1086/191268},
       adsurl = {https://ui.adsabs.harvard.edu/abs/1988ApJS...67...93C}
}

@ARTICLE{gar87,
       author = {{Garay}, Guido},
        title = "{The Orion radio zoo: PIGS (partially ionized globules), DEERS (deeply embedded energetic radio sources) and FOXES (fluctuating optical and X-ray emitting sources).}",
      journal = {\rmxaa},
         year = 1987,
        month = may,
       volume = {14},
        pages = {489-505},
       adsurl = {https://ui.adsabs.harvard.edu/abs/1987RMxAA..14..489G}
}

@INCOLLECTION{gar89,
       author = {{Garay}, G.},
        title = "{The Trapezium Radio Cluster of the Orion Nebula}",
    booktitle = {IAU Colloq. 120: Structure and Dynamics of the Interstellar Medium},
         year = 1989,
       editor = {{Tenorio-Tagle}, Guillermo and {Moles}, Mariano and {Melnick}, Jorge},
       publisher = {Springer-Verlag Berlin Heidelberg New York.},
       volume = {350},
        pages = {333-338},
          doi = {10.1007/BFb0114897},
       adsurl = {https://ui.adsabs.harvard.edu/abs/1989LNP...350..333G}
}

@INPROCEEDINGS{cos08,
       author = {{Costero}, R. and {Allen}, C. and {Echevarr{\'\i}a}, J. and {Georgiev}, L. and {Poveda}, A. and {Richer}, M.~G.},
        title = "{The Escaping Spectroscopic Binary {\ensuremath{\theta}}\^1 Ori E}",
    booktitle = {Revista Mexicana de Astronomia y Astrofisica Conference Series, vol 34,},
         year = 2008,
       series = {Revista Mexicana de Astronomia y Astrofisica Conference Series, vol 34,},
       volume = {34},
        month = dec,
        pages = {102-105},
       adsurl = {https://ui.adsabs.harvard.edu/abs/2008RMxAC..34..102C}
}

@ARTICLE{cas04,
       author = {{Castelli}, F. and {Kurucz}, R.~L.},
        title = "{Is missing Fe I opacity in stellar atmospheres a significant problem?}",
      journal = {\aap},
         year = 2004,
        month = may,
       volume = {419},
        pages = {725-733},
          doi = {10.1051/0004-6361:20040079},
       adsurl = {https://ui.adsabs.harvard.edu/abs/2004A&A...419..725C}
}

@ARTICLE{cos21,
       author = {{Costero}, Rafael and {Allen}, Christine and {Ruelas-Mayorga}, Alex and {S{\'a}nchez}, Leonardo and {Ram{\'\i}rez V{\'e}lez}, Julio and {Echevarr{\'\i}a}, Juan and {Melgoza}, Gustavo C.},
        title = "{{\'E}chelle spectroscopy of the chemically peculiar star {\ensuremath{\theta}}$^{1}$ Ori F}",
      journal = {\mnras},
         year = 2021,
        month = nov,
       volume = {507},
       number = {3},
        pages = {3400-3411},
          doi = {10.1093/mnras/stab2360},
archivePrefix = {arXiv},
       eprint = {2108.05503},
 primaryClass = {astro-ph.SR},
       adsurl = {https://ui.adsabs.harvard.edu/abs/2021MNRAS.507.3400C}
}

@ARTICLE{Rossiter1924,
       author = {{Rossiter}, R.~A.},
        title = "{On the detection of an effect of rotation during eclipse in the velocity of the brighter component of beta Lyrae, and on the constancy of velocity of this system.}",
      journal = {\apj},
         year = 1924,
        month = jul,
       volume = {60},
        pages = {15-21},
          doi = {10.1086/142825},
       adsurl = {https://ui.adsabs.harvard.edu/abs/1924ApJ....60...15R}
}

@ARTICLE{McLaughlin1924,
       author = {{McLaughlin}, D.~B.},
        title = "{Some results of a spectrographic study of the Algol system.}",
      journal = {\apj},
         year = 1924,
        month = jul,
       volume = {60},
        pages = {22-31},
          doi = {10.1086/142826},
       adsurl = {https://ui.adsabs.harvard.edu/abs/1924ApJ....60...22M}
}

@ARTICLE{Prsa2005,
       author = {{Pr{\v{s}}a}, A. and {Zwitter}, T.},
        title = "{A Computational Guide to Physics of Eclipsing Binaries. I. Demonstrations and Perspectives}",
      journal = {\apj},
         year = 2005,
        month = jul,
       volume = {628},
       number = {1},
        pages = {426-438},
          doi = {10.1086/430591},
archivePrefix = {arXiv},
       eprint = {astro-ph/0503361},
 primaryClass = {astro-ph},
       adsurl = {https://ui.adsabs.harvard.edu/abs/2005ApJ...628..426P}
}

@ARTICLE{GomezMaqueoChew2012,
       author = {{G{\'o}mez Maqueo Chew}, Yilen and {Stassun}, Keivan G. and {Pr{\v{s}}a}, Andrej and {Stempels}, Eric and {Hebb}, Leslie and {Barnes}, Rory and {Heller}, Ren{\'e} and {Mathieu}, Robert D.},
        title = "{Luminosity Discrepancy in the Equal-mass, Pre-main-sequence Eclipsing Binary Par 1802: Non-coevality or Tidal Heating?}",
      journal = {\apj},
         year = 2012,
        month = jan,
       volume = {745},
       number = {1},
          eid = {58},
        pages = {58},
          doi = {10.1088/0004-637X/745/1/58},
archivePrefix = {arXiv},
       eprint = {1111.2322},
 primaryClass = {astro-ph.SR},
       adsurl = {https://ui.adsabs.harvard.edu/abs/2012ApJ...745...58G}
}

@ARTICLE{SBS2004,
       author = {{Johnson}, D.~O.},
        title = "{Spectroscopic Binary Solver}",
      journal = {Journal of Astronomical Data},
         year = 2004,
        month = dec,
       volume = {10},
        pages = {3},
       adsurl = {https://ui.adsabs.harvard.edu/abs/2004JAD....10....3J}
}

@ARTICLE{Eisner2018,
       author = {{Eisner}, J.~A. and {Arce}, H.~G. and {Ballering}, N.~P. and {Bally}, J. and {Andrews}, S.~M. and {Boyden}, R.~D. and {Di Francesco}, J. and {Fang}, M. and {Johnstone}, D. and {Kim}, J.~S. and {Mann}, R.~K. and {Matthews}, B. and {Pascucci}, I. and {Ricci}, L. and {Sheehan}, P.~D. and {Williams}, J.~P.},
        title = "{Protoplanetary Disk Properties in the Orion Nebula Cluster: Initial Results from Deep, High-resolution ALMA Observations}",
      journal = {\apj},
         year = 2018,
        month = jun,
       volume = {860},
       number = {1},
          eid = {77},
        pages = {77},
          doi = {10.3847/1538-4357/aac3e2},
archivePrefix = {arXiv},
       eprint = {1805.03669},
 primaryClass = {astro-ph.SR},
       adsurl = {https://ui.adsabs.harvard.edu/abs/2018ApJ...860...77E}
}

@ARTICLE{Luhman2000,
       author = {{Luhman}, K.~L. and {Rieke}, G.~H. and {Young}, Erick T. and {Cotera}, Angela S. and {Chen}, H. and {Rieke}, Marcia J. and {Schneider}, Glenn and {Thompson}, Rodger I.},
        title = "{The Initial Mass Function of Low-Mass Stars and Brown Dwarfs in Young Clusters}",
      journal = {\apj},
         year = 2000,
        month = sep,
       volume = {540},
       number = {2},
        pages = {1016-1040},
          doi = {10.1086/309365},
archivePrefix = {arXiv},
       eprint = {astro-ph/0004386},
 primaryClass = {astro-ph},
       adsurl = {https://ui.adsabs.harvard.edu/abs/2000ApJ...540.1016L}
}

@ARTICLE{Gray1994,
       author = {{Gray}, David F.},
        title = "{Spectral Line-Depth Ratios as Temperature Indicators for Cool Stars}",
      journal = {\pasp},
         year = 1994,
        month = dec,
       volume = {106},
        pages = {1248},
          doi = {10.1086/133502},
       adsurl = {https://ui.adsabs.harvard.edu/abs/1994PASP..106.1248G}
}

@BOOK{moore66,
       author = {{Moore}, Charlotte E. and {Minnaert}, M.~G.~J. and {Houtgast}, J.},
        title = "{The solar spectrum 2935 A to 8770 A}",
         year = 1966,
       adsurl = {https://ui.adsabs.harvard.edu/abs/1966sst..book.....M},
      publisher = {Provided by the SAO/NASA Astrophysics Data System}
}

@misc{NIST2024,
        author = {{Kramida}, A. and {Ranchenko}, Y. and {Reader}, J. and {NIST ASD Team}},
        title = {NIST Atomic Spectra Data},
        howpublished = {\url{https://physics.nist.gov/PhysRefData/ASD/Html/verhist.shtml}},
        year = {2024},
        note = {2025}
}

@ARTICLE{Allenetal2015,
       author = {{Allen}, Christine and {Costero}, Rafael and {Hern{\'a}ndez}, Miroslava},
        title = "{The Dynamical Future of the Mini-cluster {\ensuremath{\theta}}$^{1}$ Ori B}",
      journal = {\aj},
         year = 2015,
        month = dec,
       volume = {150},
       number = {6},
          eid = {167},
        pages = {167},
          doi = {10.1088/0004-6256/150/6/167},
       adsurl = {https://ui.adsabs.harvard.edu/abs/2015AJ....150..167A}
}

@ARTICLE{Allenetal2017,
       author = {{Allen}, Christine and {Costero}, Rafael and {Ruelas-Mayorga}, Alex and {S{\'a}nchez}, L.~J.},
        title = "{On the dynamical evolution of the Orion Trapezium}",
      journal = {\mnras},
         year = 2017,
        month = apr,
       volume = {466},
       number = {4},
        pages = {4937-4953},
          doi = {10.1093/mnras/stx076},
archivePrefix = {arXiv},
       eprint = {1701.03440},
 primaryClass = {astro-ph.SR},
       adsurl = {https://ui.adsabs.harvard.edu/abs/2017MNRAS.466.4937A}
}

@ARTICLE{Lucy1971,
	   author = {{Lucy}, L.~B. and {Sweeney}, M.~A.},
	       title = "{Spectroscopic binaries with circular orbits.}",
	         journal = {\aj},
		      year = 1971,
		          month = aug,
			     volume = 76,
			         pages = {544-556},
				       doi = {10.1086/111159},
				          adsurl = {http://adsabs.harvard.edu/abs/1971AJ.....76..544L}
}

@ARTICLE{Palla1999,
       author = {{Palla}, Francesco and {Stahler}, Steven W.},
        title = "{Star Formation in the Orion Nebula Cluster}",
      journal = {\apj},
         year = 1999,
        month = nov,
       volume = {525},
       number = {2},
        pages = {772-783},
          doi = {10.1086/307928},
       adsurl = {https://ui.adsabs.harvard.edu/abs/1999ApJ...525..772P}
}

@ARTICLE{PflammKroupa2006,
       author = {{Pflamm-Altenburg}, J. and {Kroupa}, P.},
        title = "{A highly abnormal massive star mass function in the Orion Nebula cluster and the dynamical decay of trapezium systems}",
      journal = {\mnras},
         year = 2006,
        month = nov,
       volume = {373},
       number = {1},
        pages = {295-304},
          doi = {10.1111/j.1365-2966.2006.11028.x},
archivePrefix = {arXiv},
       eprint = {astro-ph/0610230},
 primaryClass = {astro-ph},
       adsurl = {https://ui.adsabs.harvard.edu/abs/2006MNRAS.373..295P}
}

@ARTICLE{Costero2019,
       author = {{Costero}, R.},
        title = "{Multiplicity of the Orion Trapezium stars}",
      journal = {arXiv e-prints},
         year = 2019,
        month = jun,
          eid = {arXiv:1906.11956},
        pages = {arXiv:1906.11956},
          doi = {10.48550/arXiv.1906.11956},
archivePrefix = {arXiv},
       eprint = {1906.11956},
 primaryClass = {astro-ph.SR},
       adsurl = {https://ui.adsabs.harvard.edu/abs/2019arXiv190611956C}
}

@book{Valle2011,
  author    = {Valle-Lira, J. A.},
  title     = {Estudio espectroscópico del Sistema Estelar V1016},
  publisher    = {UNAM},
  year = {2011},

}

@ARTICLE{Vitrichenkoetal1998,
       author = {{Vitrichenko}, E.~A. and {Klochkova}, V.~G. and {Plachinda}, S.~I.},
        title = "{A Study of the Radial Velocity of V1016 Ori}",
      journal = {Astronomy Letters},
         year = 1998,
        month = may,
       volume = {24},
        pages = {352},
       adsurl = {https://ui.adsabs.harvard.edu/abs/1998AstL...24..296V}
}

@ARTICLE{SticklandLloyd2000,
       author = {{Stickland}, D.~J. and {Lloyd}, C.},
        title = "{The strange case of theta 1 Orionis}",
      journal = {The Observatory},
         year = 2000,
        month = apr,
       volume = {120},
        pages = {141-149},
       adsurl = {https://ui.adsabs.harvard.edu/abs/2000Obs...120..141S}
}

@ARTICLE{Kounkeletal2017,
       author = {{Kounkel}, Marina and {Hartmann}, Lee and {Loinard}, Laurent and {Ortiz-Le{\'o}n}, Gisela N. and {Mioduszewski}, Amy J. and {Rodr{\'\i}guez}, Luis F. and {Dzib}, Sergio A. and {Torres}, Rosa M. and {Pech}, Gerardo and {Galli}, Phillip A.~B. and {Rivera}, Juana L. and {Boden}, Andrew F. and {Evans}, II, Neal J. and {Brice{\~n}o}, Cesar and {Tobin}, John J.},
        title = "{The Gould{\textquoteright}s Belt Distances Survey (GOBELINS) II. Distances and Structure toward the Orion Molecular Clouds}",
      journal = {\apj},
         year = 2017,
        month = jan,
       volume = {834},
       number = {2},
          eid = {142},
        pages = {142},
          doi = {10.3847/1538-4357/834/2/142},
archivePrefix = {arXiv},
       eprint = {1609.04041},
 primaryClass = {astro-ph.SR},
}

@ARTICLE{Dzibetal2021,
       author = {{Dzib}, Sergio A. and {Forbrich}, Jan and {Reid}, Mark J. and {Menten}, Karl M.},
        title = "{A VLBA Survey of Radio Stars in the Orion Nebula Cluster. II. Astrometry}",
      journal = {\apj},
         year = 2021,
        month = jan,
       volume = {906},
       number = {1},
          eid = {24},
        pages = {24},
          doi = {10.3847/1538-4357/abc68f},
archivePrefix = {arXiv},
       eprint = {2011.09331},
 primaryClass = {astro-ph.SR},
       adsurl = {https://ui.adsabs.harvard.edu/abs/2021ApJ...906...24D}
}

@ARTICLE{Petretal1998,
       author = {{Petr}, Monika G. and {Coud{\'e} du Foresto}, Vincent and {Beckwith}, Steven V.~W. and {Richichi}, Andrea and {McCaughrean}, Mark J.},
        title = "{Binary Stars in the Orion Trapezium Cluster Core}",
      journal = {\apj},
         year = 1998,
        month = jun,
       volume = {500},
       number = {2},
        pages = {825-837},
          doi = {10.1086/305751},
       adsurl = {https://ui.adsabs.harvard.edu/abs/1998ApJ...500..825P}
}

@ARTICLE{Karletal2018,
       author = {{GRAVITY Collaboration} and {Karl}, Martina and {Pfuhl}, Oliver and {Eisenhauer}, Frank and {Genzel}, Reinhard and {Grellmann}, Rebekka and {Habibi}, Maryam and {Abuter}, Roberto and {Accardo}, Matteo and {Amorim}, Ant{\'o}nio and {Anugu}, Narsireddy and {{\'A}vila}, Gerardo and {Benisty}, Myriam and {Berger}, Jean-Philippe and {Blind}, Nicolas and {Bonnet}, Henri and {Bourget}, Pierre and {Brandner}, Wolfgang and {Brast}, Roland and {Buron}, Alexander and {Caratti O Garatti}, Alessio and {Chapron}, Fr{\'e}d{\'e}ric and {Cl{\'e}net}, Yann and {Collin}, Claude and {Coud{\'e} Du Foresto}, Vincent and {de Wit}, Willem-Jan and {de Zeeuw}, Tim and {Deen}, Casey and {Delplancke-Str{\"o}bele}, Fran{\c{c}}oise and {Dembet}, Roderick and {Derie}, Fr{\'e}d{\'e}ric and {Dexter}, Jason and {Duvert}, Gilles and {Ebert}, Monica and {Eckart}, Andreas and {Esselborn}, Michael and {F{\'e}dou}, Pierre and {Finger}, Gert and {Garcia}, Paulo and {Garcia Dabo}, Cesar Enrique and {Garcia Lopez}, Rebeca and {Gao}, Feng and {Gendron}, {\'E}ric and {Gillessen}, Stefan and {Gont{\'e}}, Fr{\'e}d{\'e}ric and {Gordo}, Paulo and {Gr{\"o}zinger}, Ulrich and {Guajardo}, Patricia and {Guieu}, Sylvain and {Haguenauer}, Pierre and {Hans}, Oliver and {Haubois}, Xavier and {Haug}, Marcus and {Hau{\ss}mann}, Frank and {Henning}, Thomas and {Hippler}, Stefan and {Horrobin}, Matthew and {Huber}, Armin and {Hubert}, Zoltan and {Hubin}, Norbert and {Jakob}, Gerd and {Jochum}, Lieselotte and {Jocou}, Laurent and {Kaufer}, Andreas and {Kellner}, Stefan and {Kendrew}, Sarah and {Kern}, Lothar and {Kervella}, Pierre and {Kiekebusch}, Mario and {Klein}, Ralf and {K{\"o}hler}, Rainer and {Kolb}, Johan and {Kulas}, Martin and {Lacour}, Sylvestre and {Lapeyr{\`e}re}, Vincent and {Lazareff}, Bernard and {Le Bouquin}, Jean-Baptiste and {L{\'e}na}, Pierre and {Lenzen}, Rainer and {L{\'e}v{\^e}que}, Samuel and {Lin}, Chien-Cheng and {Lippa}, Magdalena and {Magnard}, Yves and {Mehrgan}, Leander and {M{\'e}rand}, Antoine and {Moulin}, Thibaut and {M{\"u}ller}, Eric and {M{\"u}ller}, Friedrich and {Neumann}, Udo and {Oberti}, Sylvain and {Ott}, Thomas and {Pallanca}, Laurent and {Panduro}, Johana and {Pasquini}, Luca and {Paumard}, Thibaut and {Percheron}, Isabelle and {Perraut}, Karine and {Perrin}, Guy and {Pfl{\"u}ger}, Andreas and {Duc}, Thanh Phan and {Plewa}, Philipp M. and {Popovic}, Dan and {Rabien}, Sebastian and {Ram{\'\i}rez}, Andr{\'e}s and {Ramos}, Jose and {Rau}, Christian and {Riquelme}, Miguel and {Rodr{\'\i}guez-Coira}, Gustavo and {Rohloff}, Ralf-Rainer and {Rosales}, Alejandra and {Rousset}, G{\'e}rard and {Sanchez-Bermudez}, Joel and {Scheithauer}, Silvia and {Sch{\"o}ller}, Markus and {Schuhler}, Nicolas and {Spyromilio}, Jason and {Straub}, Odele and {Straubmeier}, Christian and {Sturm}, Eckhard and {Suarez}, Marcos and {Tristram}, Konrad R.~W. and {Ventura}, Noel and {Vincent}, Fr{\'e}d{\'e}ric and {Waisberg}, Idel and {Wank}, Imke and {Widmann}, Felix and {Wieprecht}, Ekkehard and {Wiest}, Michael and {Wiezorrek}, Erich and {Wittkowski}, Markus and {Woillez}, Julien and {Wolff}, Burkhard and {Yazici}, Senol and {Ziegler}, Denis and {Zins}, G{\'e}rard},
        title = "{Multiple star systems in the Orion nebula}",
      journal = {\aap},
         year = 2018,
        month = dec,
       volume = {620},
          eid = {A116},
        pages = {A116},
          doi = {10.1051/0004-6361/201833575},
archivePrefix = {arXiv},
       eprint = {1809.10376},
 primaryClass = {astro-ph.SR},
       adsurl = {https://ui.adsabs.harvard.edu/abs/2018A&A...620A.116G}
}

@ARTICLE{Maizetal2022,
       author = {{Ma{\'\i}z Apell{\'a}niz}, J. and {Barb{\'a}}, R.~H. and {Fern{\'a}ndez Aranda}, R. and {Pantaleoni Gonz{\'a}lez}, M. and {Crespo Bellido}, P. and {Sota}, A. and {Alfaro}, E.~J.},
        title = "{The Villafranca catalog of Galactic OB groups. II. From Gaia DR2 to EDR3 and ten new systems with O stars}",
      journal = {\aap},
         year = 2022,
        month = jan,
       volume = {657},
          eid = {A131},
        pages = {A131},
          doi = {10.1051/0004-6361/202142364},
archivePrefix = {arXiv},
       eprint = {2110.01464},
 primaryClass = {astro-ph.GA},
       adsurl = {https://ui.adsabs.harvard.edu/abs/2022A&A...657A.131M}
}

@ARTICLE{Weietal2024,
       author = {{Wei}, Lingfeng and {Theissen}, Christopher A. and {Konopacky}, Quinn M. and {Lu}, Jessica R. and {Hsu}, Chih-Chun and {Kim}, Dongwon},
        title = "{The 3D Kinematics of the Orion Nebula Cluster. II. Mass-dependent Kinematics of the Inner Cluster}",
      journal = {\apj},
         year = 2024,
        month = feb,
       volume = {962},
       number = {2},
          eid = {174},
        pages = {174},
          doi = {10.3847/1538-4357/ad1401},
archivePrefix = {arXiv},
       eprint = {2312.04751},
 primaryClass = {astro-ph.GA},
       adsurl = {https://ui.adsabs.harvard.edu/abs/2024ApJ...962..174W}
}

@ARTICLE{shulz24,
       author = {{Schulz}, Norbert S. and {Huenemoerder}, David P. and {Principe}, David A. and {Gagne}, Marc and {G{\"u}nther}, Hans Moritz and {Kastner}, Joel and {Nichols}, Joy and {Pollock}, Andrew and {Preibisch}, Thomas and {Testa}, Paola and {Reale}, Fabio and {Favata}, Fabio and {Canizares}, Claude R.},
        title = "{The Nature of X-Rays from Young Stellar Objects in the Orion Nebula Cluster{\textemdash}A Chandra HETGS Legacy Project}",
      journal = {\apj},
         year = 2024,
        month = aug,
       volume = {970},
       number = {2},
          eid = {190},
        pages = {190},
          doi = {10.3847/1538-4357/ad47c2},
archivePrefix = {arXiv},
       eprint = {2404.19676},
 primaryClass = {astro-ph.SR},
       adsurl = {https://ui.adsabs.harvard.edu/abs/2024ApJ...970..190S}
}

@ARTICLE{Theisenetal2022,
       author = {{Theissen}, Christopher A. and {Konopacky}, Quinn M. and {Lu}, Jessica R. and {Kim}, Dongwon and {Zhang}, Stella Y. and {Hsu}, Chih-Chun and {Chu}, Laurie and {Wei}, Lingfeng},
        title = "{The 3D Kinematics of the Orion Nebula Cluster: NIRSPEC-AO Radial Velocities of the Core Population}",
      journal = {\apj},
         year = 2022,
        month = feb,
       volume = {926},
       number = {2},
          eid = {141},
        pages = {141},
          doi = {10.3847/1538-4357/ac3252},
archivePrefix = {arXiv},
       eprint = {2105.05871},
 primaryClass = {astro-ph.GA},
       adsurl = {https://ui.adsabs.harvard.edu/abs/2022ApJ...926..141T}
}

@ARTICLE{maiz21,
       author = {{Ma{\'\i}z Apell{\'a}niz}, J. and {Pantaleoni Gonz{\'a}lez}, M. and {Barb{\'a}}, R.~H.},
        title = "{{\ensuremath{\theta}}$^{1}$ Ori C as a Medieval Bully: A Possible Very Recent Ejection in the Trapezium}",
      journal = {Research Notes of the American Astronomical Society},
         year = 2021,
        month = oct,
       volume = {5},
       number = {10},
          eid = {232},
        pages = {232},
          doi = {10.3847/2515-5172/ac2eee},
archivePrefix = {arXiv},
       eprint = {2110.08022},
 primaryClass = {astro-ph.SR},
       adsurl = {https://ui.adsabs.harvard.edu/abs/2021RNAAS...5..232M}
}

@ARTICLE{Gaia2016,
       author = {{Gaia Collaboration} and {Prusti}, T. and {de Bruijne}, J.~H.~J. and {Brown}, A.~G.~A. and {Vallenari}, A. and {Babusiaux}, C. and {Bailer-Jones}, C.~A.~L. and {Bastian}, U. and {Biermann}, M. and {Evans}, D.~W. and {Eyer}, L. and {Jansen}, F. and {Jordi}, C. and {Klioner}, S.~A. and {Lammers}, U. and {Lindegren}, L. and {Luri}, X. and {Mignard}, F. and {Milligan}, D.~J. and {Panem}, C. and {Poinsignon}, V. and {Pourbaix}, D. and {Randich}, S. and {Sarri}, G. and {Sartoretti}, P. and {Siddiqui}, H.~I. and {Soubiran}, C. and {Valette}, V. and {van Leeuwen}, F. and {Walton}, N.~A. and {Aerts}, C. and {Arenou}, F. and {Cropper}, M. and {Drimmel}, R. and {H{\o}g}, E. and {Katz}, D. and {Lattanzi}, M.~G. and {O'Mullane}, W. and {Grebel}, E.~K. and {Holland}, A.~D. and {Huc}, C. and {Passot}, X. and {Bramante}, L. and {Cacciari}, C. and {Casta{\~n}eda}, J. and {Chaoul}, L. and {Cheek}, N. and {De Angeli}, F. and {Fabricius}, C. and {Guerra}, R. and {Hern{\'a}ndez}, J. and {Jean-Antoine-Piccolo}, A. and {Masana}, E. and {Messineo}, R. and {Mowlavi}, N. and {Nienartowicz}, K. and {Ord{\'o}{\~n}ez-Blanco}, D. and {Panuzzo}, P. and {Portell}, J. and {Richards}, P.~J. and {Riello}, M. and {Seabroke}, G.~M. and {Tanga}, P. and {Th{\'e}venin}, F. and {Torra}, J. and {Els}, S.~G. and {Gracia-Abril}, G. and {Comoretto}, G. and {Garcia-Reinaldos}, M. and {Lock}, T. and {Mercier}, E. and {Altmann}, M. and {Andrae}, R. and {Astraatmadja}, T.~L. and {Bellas-Velidis}, I. and {Benson}, K. and {Berthier}, J. and {Blomme}, R. and {Busso}, G. and {Carry}, B. and {Cellino}, A. and {Clementini}, G. and {Cowell}, S. and {Creevey}, O. and {Cuypers}, J. and {Davidson}, M. and {De Ridder}, J. and {de Torres}, A. and {Delchambre}, L. and {Dell'Oro}, A. and {Ducourant}, C. and {Fr{\'e}mat}, Y. and {Garc{\'\i}a-Torres}, M. and {Gosset}, E. and {Halbwachs}, J. -L. and {Hambly}, N.~C. and {Harrison}, D.~L. and {Hauser}, M. and {Hestroffer}, D. and {Hodgkin}, S.~T. and {Huckle}, H.~E. and {Hutton}, A. and {Jasniewicz}, G. and {Jordan}, S. and {Kontizas}, M. and {Korn}, A.~J. and {Lanzafame}, A.~C. and {Manteiga}, M. and {Moitinho}, A. and {Muinonen}, K. and {Osinde}, J. and {Pancino}, E. and {Pauwels}, T. and {Petit}, J. -M. and {Recio-Blanco}, A. and {Robin}, A.~C. and {Sarro}, L.~M. and {Siopis}, C. and {Smith}, M. and {Smith}, K.~W. and {Sozzetti}, A. and {Thuillot}, W. and {van Reeven}, W. and {Viala}, Y. and {Abbas}, U. and {Abreu Aramburu}, A. and {Accart}, S. and {Aguado}, J.~J. and {Allan}, P.~M. and {Allasia}, W. and {Altavilla}, G. and {{\'A}lvarez}, M.~A. and {Alves}, J. and {Anderson}, R.~I. and {Andrei}, A.~H. and {Anglada Varela}, E. and {Antiche}, E. and {Antoja}, T. and {Ant{\'o}n}, S. and {Arcay}, B. and {Atzei}, A. and {Ayache}, L. and {Bach}, N. and {Baker}, S.~G. and {Balaguer-N{\'u}{\~n}ez}, L. and {Barache}, C. and {Barata}, C. and {Barbier}, A. and {Barblan}, F. and {Baroni}, M. and {Barrado y Navascu{\'e}s}, D. and {Barros}, M. and {Barstow}, M.~A. and {Becciani}, U. and {Bellazzini}, M. and {Bellei}, G. and {Bello Garc{\'\i}a}, A. and {Belokurov}, V. and {Bendjoya}, P. and {Berihuete}, A. and {Bianchi}, L. and {Bienaym{\'e}}, O. and {Billebaud}, F. and {Blagorodnova}, N. and {Blanco-Cuaresma}, S. and {Boch}, T. and {Bombrun}, A. and {Borrachero}, R. and {Bouquillon}, S. and {Bourda}, G. and {Bouy}, H. and {Bragaglia}, A. and {Breddels}, M.~A. and {Brouillet}, N. and {Br{\"u}semeister}, T. and {Bucciarelli}, B. and {Budnik}, F. and {Burgess}, P. and {Burgon}, R. and {Burlacu}, A. and {Busonero}, D. and {Buzzi}, R. and {Caffau}, E. and {Cambras}, J. and {Campbell}, H. and {Cancelliere}, R. and {Cantat-Gaudin}, T. and {Carlucci}, T. and {Carrasco}, J.~M. and {Castellani}, M. and {Charlot}, P. and {Charnas}, J. and {Charvet}, P. and {Chassat}, F. and {Chiavassa}, A. and {Clotet}, M. and {Cocozza}, G. and {Collins}, R.~S. and {Collins}, P. and {Costigan}, G.},
        title = "{The Gaia mission}",
      journal = {\aap},
         year = 2016,
        month = nov,
       volume = {595},
          eid = {A1},
        pages = {A1},
          doi = {10.1051/0004-6361/201629272},
archivePrefix = {arXiv},
       eprint = {1609.04153},
 primaryClass = {astro-ph.IM},
       adsurl = {https://ui.adsabs.harvard.edu/abs/2016A&A...595A...1G}
}

@ARTICLE{GaiaDR32023,
       author = {{Gaia Collaboration} and {Vallenari}, A. and {Brown}, A.~G.~A. and {Prusti}, T. and {de Bruijne}, J.~H.~J. and {Arenou}, F. and {Babusiaux}, C. and {Biermann}, M. and {Creevey}, O.~L. and {Ducourant}, C. and {Evans}, D.~W. and {Eyer}, L. and {Guerra}, R. and {Hutton}, A. and {Jordi}, C. and {Klioner}, S.~A. and {Lammers}, U.~L. and {Lindegren}, L. and {Luri}, X. and {Mignard}, F. and {Panem}, C. and {Pourbaix}, D. and {Randich}, S. and {Sartoretti}, P. and {Soubiran}, C. and {Tanga}, P. and {Walton}, N.~A. and {Bailer-Jones}, C.~A.~L. and {Bastian}, U. and {Drimmel}, R. and {Jansen}, F. and {Katz}, D. and {Lattanzi}, M.~G. and {van Leeuwen}, F. and {Bakker}, J. and {Cacciari}, C. and {Casta{\~n}eda}, J. and {De Angeli}, F. and {Fabricius}, C. and {Fouesneau}, M. and {Fr{\'e}mat}, Y. and {Galluccio}, L. and {Guerrier}, A. and {Heiter}, U. and {Masana}, E. and {Messineo}, R. and {Mowlavi}, N. and {Nicolas}, C. and {Nienartowicz}, K. and {Pailler}, F. and {Panuzzo}, P. and {Riclet}, F. and {Roux}, W. and {Seabroke}, G.~M. and {Sordo}, R. and {Th{\'e}venin}, F. and {Gracia-Abril}, G. and {Portell}, J. and {Teyssier}, D. and {Altmann}, M. and {Andrae}, R. and {Audard}, M. and {Bellas-Velidis}, I. and {Benson}, K. and {Berthier}, J. and {Blomme}, R. and {Burgess}, P.~W. and {Busonero}, D. and {Busso}, G. and {C{\'a}novas}, H. and {Carry}, B. and {Cellino}, A. and {Cheek}, N. and {Clementini}, G. and {Damerdji}, Y. and {Davidson}, M. and {de Teodoro}, P. and {Nu{\~n}ez Campos}, M. and {Delchambre}, L. and {Dell'Oro}, A. and {Esquej}, P. and {Fern{\'a}ndez-Hern{\'a}ndez}, J. and {Fraile}, E. and {Garabato}, D. and {Garc{\'\i}a-Lario}, P. and {Gosset}, E. and {Haigron}, R. and {Halbwachs}, J. -L. and {Hambly}, N.~C. and {Harrison}, D.~L. and {Hern{\'a}ndez}, J. and {Hestroffer}, D. and {Hodgkin}, S.~T. and {Holl}, B. and {Jan{\ss}en}, K. and {Jevardat de Fombelle}, G. and {Jordan}, S. and {Krone-Martins}, A. and {Lanzafame}, A.~C. and {L{\"o}ffler}, W. and {Marchal}, O. and {Marrese}, P.~M. and {Moitinho}, A. and {Muinonen}, K. and {Osborne}, P. and {Pancino}, E. and {Pauwels}, T. and {Recio-Blanco}, A. and {Reyl{\'e}}, C. and {Riello}, M. and {Rimoldini}, L. and {Roegiers}, T. and {Rybizki}, J. and {Sarro}, L.~M. and {Siopis}, C. and {Smith}, M. and {Sozzetti}, A. and {Utrilla}, E. and {van Leeuwen}, M. and {Abbas}, U. and {{\'A}brah{\'a}m}, P. and {Abreu Aramburu}, A. and {Aerts}, C. and {Aguado}, J.~J. and {Ajaj}, M. and {Aldea-Montero}, F. and {Altavilla}, G. and {{\'A}lvarez}, M.~A. and {Alves}, J. and {Anders}, F. and {Anderson}, R.~I. and {Anglada Varela}, E. and {Antoja}, T. and {Baines}, D. and {Baker}, S.~G. and {Balaguer-N{\'u}{\~n}ez}, L. and {Balbinot}, E. and {Balog}, Z. and {Barache}, C. and {Barbato}, D. and {Barros}, M. and {Barstow}, M.~A. and {Bartolom{\'e}}, S. and {Bassilana}, J. -L. and {Bauchet}, N. and {Becciani}, U. and {Bellazzini}, M. and {Berihuete}, A. and {Bernet}, M. and {Bertone}, S. and {Bianchi}, L. and {Binnenfeld}, A. and {Blanco-Cuaresma}, S. and {Blazere}, A. and {Boch}, T. and {Bombrun}, A. and {Bossini}, D. and {Bouquillon}, S. and {Bragaglia}, A. and {Bramante}, L. and {Breedt}, E. and {Bressan}, A. and {Brouillet}, N. and {Brugaletta}, E. and {Bucciarelli}, B. and {Burlacu}, A. and {Butkevich}, A.~G. and {Buzzi}, R. and {Caffau}, E. and {Cancelliere}, R. and {Cantat-Gaudin}, T. and {Carballo}, R. and {Carlucci}, T. and {Carnerero}, M.~I. and {Carrasco}, J.~M. and {Casamiquela}, L. and {Castellani}, M. and {Castro-Ginard}, A. and {Chaoul}, L. and {Charlot}, P. and {Chemin}, L. and {Chiaramida}, V. and {Chiavassa}, A. and {Chornay}, N. and {Comoretto}, G. and {Contursi}, G. and {Cooper}, W.~J. and {Cornez}, T. and {Cowell}, S. and {Crifo}, F. and {Cropper}, M. and {Crosta}, M. and {Crowley}, C. and {Dafonte}, C. and {Dapergolas}, A. and {David}, M. and {David}, P. and {de Laverny}, P. and {De Luise}, F. and {De March}, R.},
        title = "{Gaia Data Release 3. Summary of the content and survey properties}",
      journal = {\aap},
         year = 2023,
        month = jun,
       volume = {674},
          eid = {A1},
        pages = {A1},
          doi = {10.1051/0004-6361/202243940},
archivePrefix = {arXiv},
       eprint = {2208.00211},
 primaryClass = {astro-ph.GA},
       adsurl = {https://ui.adsabs.harvard.edu/abs/2023A&A...674A...1G}
}

\end{document}